%% file: qrf.tex
\documentclass[11pt]{article}
\input{packages.tex}

\input{standard.tex}

\input{Alg_Macro}

\usepackage{float}

\usepackage{tikz-cd}
\newcommand{\thistitle}{Quantum Reference Frames from Top-Down Crossed Products} 
 
\begin{document}

\title{\thistitle}
\author{
Shadi Ali Ahmad$^{a},$ Wissam Chemissany$^{b},$ Marc S. Klinger$^{c}$, and Robert G. Leigh$^{c}$
	\\
	\\
    {\small \emph{\nyu{a}}} \\ \\
    {\small \emph{\upenn{b}}} \\ \\
	{\small \emph{\uiuc{c}}}
	\\
	}
\date{}
\maketitle
\vspace{-0.5cm}
\begin{abstract}
All physical observations are made relative to a reference frame, which is a system in its own right. If the system of interest admits a group symmetry, the reference frame observing it must transform commensurately under the group to ensure the covariance of the combined system. We point out that the crossed product is a way to realize quantum reference frames from the \textit{bottom-up}; adjoining a quantum reference frame and imposing constraints generates a crossed product algebra. We provide a top-down specification of crossed product algebras and show that one cannot obtain inequivalent quantum reference frames using this approach. As a remedy, we define an abstract algebra associated to the system and symmetry group built out of \textit{relational} crossed product algebras associated with different choices of quantum reference frames. We term this object the G-framed algebra, and show how potentially inequivalent frames are realized within this object. We comment on this algebra's analog of the classical Gribov problem in gauge theory, its importance in gravity where we show that it is relevant for semiclassical de Sitter and potentially beyond the semiclassical limit, and its utility for understanding the frame-dependence of physical notions like observables, density states, and entropies.
\vspace{0.3cm}
\end{abstract}
\begingroup
\hypersetup{linkcolor=black}
\tableofcontents
\endgroup
\noindent\rule{\textwidth}{0.6pt}
\setcounter{footnote}{0}
\renewcommand{\thefootnote}{\arabic{footnote}}

\newcommand{\curr}[1]{\mathbb{J}_{#1}}
\newcommand{\constr}[1]{\mathbb{M}_{#1}}
\newcommand{\chgdens}[1]{\mathbb{Q}_{#1}}
\newcommand{\spac}[1]{S_{#1}}
\newcommand{\hyper}[1]{\Sigma_{#1}}
\newcommand{\chg}[1]{\mathbb{H}_{#1}}
\newcommand{\ThomSig}[1]{\hat{\Sigma}_{#1}}
\newcommand{\ThomS}[1]{\hat{S}_{#1}}
\newcommand{\discuss}[1]{{\color{red} #1}}
\newtheorem{theorem}{Theorem}
\newtheorem{lemma}[equation]{Lemma}
\newtheorem{proposition}[equation]{Proposition}
\newtheorem{corollary}[equation]{Corollary}
\newtheorem{definition}{Definition}
\newtheorem{construction}[equation]{Construction}
\newtheorem{question}[equation]{Question}
\newtheorem{convention}[equation]{Convention}
\newtheorem{problem}[equation]{Problem}
\newtheorem{example}[equation]{Example}
\newtheorem{exercise}[equation]{Exercise}
\newtheorem{remark}[equation]{Remark}
\newtheorem{physics}[equation]{Physics}
\newtheorem{notation}[equation]{Notation}
\newtheorem{noterm}[equation]{Notation and Terminology}
\newtheorem{conj}[equation]{Conjecture}

\newcommand{\gravgroup}{G_L}
\newcommand{\gaugegroup}{G_G}
\newcommand{\gravL}{L_L}
\newcommand{\gaugeL}{L_G} 
\newcommand{\solder}{e}
\newcommand{\sym}{\Omega}
\newcommand{\symdens}{\omega}
\newcommand{\sympot}{\Theta}
\newcommand{\sympotdens}{\theta}
\newcommand{\chgcon}[1]{\chg{#1}^{(1)}}
\newcommand{\chgchg}[1]{\chg{#1}^{(2)}}
\newcommand{\cpalg}{\mathcal{M}_{cp}}

\newcommand{\red}[1]{{\color{red} #1}}
\newcommand{\purple}[1]{{\color{purple} #1}}

\section{Introduction}
A central problem in physics is understanding systems with constraints. The paradigmatic example of such a system is a gauge theory in which constraints are regarded as encoding redundancies of description relative to local symmetries. Constraints, whether due to symmetry or otherwise, imbue a theory with interesting global structures which complicate the resulting classical and quantum descriptions. These structures have been the subject of a great deal of work in historic literature \cite{dirac1964lectures,dirac_lectures_2013,Ashtekar:1995zh,Marolf:1996gb,Giulini:1998kf,Giulini:1998rk,Marolf:2000iq} often going under the name of constraint quantization. 

One such theory with constraints is gravity with its diffeomorphism invariance and background independence~\cite{Ashtekar:1995zh, Ashtekar1991-ASHCPO, rovelli_2004, thiemann_2007,Witten:2023xze}. An important lesson of general relativity is that only relational data is physical. Any measurement is done relative to a reference frame, or colloquially an observer. This observer in general has internal degrees of freedom of its own, and since it is a part of the universe that it observes, must also be subordinate to the same physical laws. Thus, one must treat the observer itself as a quantum system in the pursuit of obtaining a more coherent synthesis of the laws of nature. This conviction gives rise to the idea of a quantum reference frame (QRF), which has proliferated across different disciplines of physics~\cite{aharonov_quantum_1984, rovelli_quantum_1991, kitaev_superselection_2004, bartlett_reference_2007, gour_resource_2008,  girelli_quantum_2008, bartlett_quantum_2009, angelo_physics_2011, angelo_kinematics_2012,  palmer_changing_2014, pienaar_relational_2016, smith_quantum_2016, miyadera_approximating_2016, loveridge_relativity_2017, belenchia_quantum_2018, giacomini_quantum_2019, hoehn_equivalence_2020, castro-ruiz_quantum_2020, Danielson:2021egj,hoehn_trinity_2021, castro-ruiz_relative_2021, giacomini_spacetime_2021, ali_ahmad_quantum_2022, cepollaro_quantum_2022, giacomini_second-quantized_2022, apadula_quantum_2022}. When the system to be observed admits a symmetry, the observer must carry a representation of that symmetry in order to ensure the covariance of the total system. Constraints between the observer and system arise naturally from the perspective of QRFs and one may view QRFs as a way of implementing constraint quantization.

Recently, another way of performing constraint quantization has proven to be interesting and physically relevant, known as the crossed product\cite{witten_gravity_2022,chandrasekaran_large_2022,chandrasekaran_algebra_2022,AliAhmad:2023etg,Klinger:2023tgi,Klinger:2023auu}. The crossed product of a system's algebra of observables by its symmetry group naturally implements the constraints implied by the symmetry.\footnote{The analysis in this paper is sufficiently general to apply to any locally compact group.} This interpretation is due to a central result called the commutation theorem~\cite{van1978continuous}, which posits that one can view the construction of a crossed product algebra as a two-step process: (1) adjoining degrees of freedom transforming appropriately under the group and (2) identifying it as the fixed point subalgebra of the resulting system under the relevant action of group. Our work here begins with the observation that the crossed product is precisely an instantiation of the principles of QRFs, with a very particular choice of quantum reference frame. This is formulated directly in terms of Page-Wooters reduction maps \cite{page_evolution_1983,wootters_time_1984}. 

A natural question then arises: can one generate different quantum reference frames for the same system? We answer this question in the negative and show that while the gauge-invariant algebra of a system and any valid QRF gives rise to a crossed product algebra, one cannot obtain \textit{inequivalent} reference frames using this approach. In proving this, we expand on an alternative characterization of the crossed product that we refer to as the top-down approach which may be useful for the community~\cite{landstad1979duality}. An intuitive reason for our result is that different reference frames observe different systems even if they descend from the same kinematical space. So, to obtain an algebra which allows for multiple QRFs, one must admit the possibility of different system algebras before taking the crossed product. This is known in the literature as quantum subsystem relativity~\cite{AliAhmad:2021adn, castro-ruiz_relative_2021,delaHamette:2021oex, Hoehn:2023sub}. 

Since one can have many different frames that probe vastly different parts of the same system, we propose an algebra generalizing the idea of a crossed product which retains the possibility of admitting inequivalent choices of QRF. We term this object the G-framed algebra, and demonstrate how it may be regarded as an algebraic analog of a global quotient space. A generic $G$-framed algebra is covered by an atlas of local crossed product charts, each coinciding with the constraint quantized physics of a given QRF. The different QRFs here do not have to be auxiliary systems arising from adjoining group degrees of freedom, and could very well be internal to the system of interest as long as they transform suitably under the group. The $G$-framed algebra allows us to make sense of frame-dependent notions like observables, density states, measurements, and entropies. Rigorously, the $G$-framed algebra can be thought of as an algebraic version of an orbifold which encodes the global quotient of a manifold by a Lie group~\cite{caramello2022introduction}. Each local crossed product chart coincides morally with the quantization of the local quotient space described by an orbifold chart, representing the gauge-invariant physics as described by the corresponding choice of QRF. The stitching together of different descriptions of the same fundamental theory is carried out by a non-trivial quotienting procedure imbuing the G-framed algebra with global and frame-independent data. 

Building upon this point of view, we motivate the need for a $G$-framed algebra in generic gauge theories by appealing to the Gribov problem. In the classical context, the Gribov problem designates that the constraint surface in a typical gauge theory will have the structure of an orbifold \cite{Langmann:1993pb,1994CMaPh.160..431M}, and thus in our language will induce multiple distinct QRFs under quantization. We also point out that there are two levels to the Gribov problem: one in which the quotient space is actually a manifold in which case the quotient of the $G$-framed algebra gets rid of the extraneous Gribov copies, and another where the quotient space is a true orbifold in which case the $G$-framed algebra will have non-trivial global structure.
\subsection{Outline}
The current paper is organized as follows. In Section \ref{sec: crossed products} we review crossed product algebras with a focus on two main points: (1) the interpretation of crossed products as fixed points of kinematical algebras with respect to group automorphisms, and (2) the relationship between crossed product algebras induced from a fixed covariant system in various covariant representations. At the end of the day, the top down approach to the crossed product (Section \ref{sec: top down}) demonstrates that all crossed product algebras induced from a common covariant system are isomorphic, and therefore possess an interpretation as the physical operators relative to a particular constraint imposed on a kinematical space. In Section \ref{sec: cp-qrf} we prove that, given a fixed QRF encoded by a covariant system, any internal reference frame induces a covariant representation. Thus, the relational crossed products induced by any pair of internal frames for a fixed QRF are isomorphic. We therefore hypothesize that QRFs are primarily useful in algebraic contexts where multiple covariant systems are mutually present, as exemplified by the tripartite algebra example (Section \ref{sec: tripartite example}). This leads us to Section \ref{sec: big crossed} in which we define the $G$-framed algebra. A $G$-framed algebra is a perspective neutral algebra of gauge invariant operators in which individual crossed product algebras are interpreted as local charts encoding the QRF of a particular observer. The $G$-framed algebra is an algebraic analog of an orbifold which is comprised of local quotient spaces and appears naturally in the classical phase space description of a constrained system in relation to the Gribov ambiguity. It is also naturally applied to semiclassical gravity in the static patch of de Sitter space with multiple observers (Section \ref{sec: de sitter}), and to the quantification of frame dependence for many gauge invariant observables (Section \ref{sec: rel states}). We conclude in Section \ref{sec: disc} with a discussion of various directions for future work centered around the $G$-framed algebra as a tool for encoding the global quantum structure of constrained systems, gauge theories, and gravity. 
\subsection{Note}
During the final stages of writing this paper, \cite{Fewster:2024pur,DeVuyst:2024pop} appeared which contain some overlap with our work. The first reference mainly dealt with the crossed product as a way of generating different frames interpreted as measurement probes for the \textit{same system algebra}. While QRFs and measurement theory are deeply intertwined, the system itself is different for different choices of frames in our work which marks a drastic difference. The second reference is more aligned with our work, however our $G$-framed algebra allows for different choices of QRFs that may not be auxiliary to one fixed system of interest shared by all frames. We hope to understand the relations between the $G$-framed algebra and the approaches of these references more in the future. Finally, we would like to mention that Refs.~\cite{Gomez:2022eui,  Gomez:2023wrq, Gomez:2023upk} were recently brought to our attention, which contain discussions on the relevance of the crossed product in the context of cosmology where clocks are viewed as quantum reference frames.

\subsection{Overview of results}

The first part of the current paper reviews the construction and interpretation of a single crossed product algebra from a bottom up and a top down point of view. From the bottom up point of view (which is the most common presentation in the literature) the construction of a crossed product algebra begins with the specification of a covariant system $(M,G,\alpha)$. Here, $M$ is a von Neumann algebra which we will refer to as the `system' algebra, $G$ is a locally compact group, and $\alpha: G \rightarrow \text{Aut}(M)$ is an action of the group $G$ on $M$. Given a covariant system $(M,G,\alpha)$ a covariant representation is a triple $(K,\pi_{\alpha}^{(K)}, \lambda^{(K)})$, where $K$ is a Hilbert space, $\pi_{\alpha}^{(K)}$ is a faithful representation of the system $M$ and $\lambda^{(K)}$ is a unitary representation of the group $G$ which implements the automorphism $\alpha$ as an adjoint action within the larger algebra $B(K)$. The crossed product induced from the covariant system $(M,G,\alpha)$ in the covariant representation $(K,\pi_{\alpha}^{(K)},\lambda^{(K)})$ is simply the von Neumann algebra generated by $\pi_{\alpha}^{(K)}(M)$ and $\lambda^{(K)}(G)$, which we denote by $M \rtimes_{\alpha}^{(K)} G$. Colloquially, one may think of this algebra as consisting of the system degrees of freedom $M$ along with the generators of the automorphism $\alpha$. Since the construction of $M \rtimes_{\alpha}^{(K)} G$ includes taking a weak closure in the topology induced by the Hilbert space $K$, it is not immediately clear whether $M \rtimes_{\alpha}^{(K)} G$ is defined independently of the chosen covariant representation. 

The bottom up point of view also implicates two other useful ways of interpreting a crossed product. The first approach follows Haagerup's presentation in which the crossed product is induced from an associated $C^*$ algebra consisting of maps from the group $G$ into the system algebra $M$ \cite{haagerup1978dualI,haagerup1978dualII}. This algebra is rendered into a von Neumann algebra by inducing a Hilbert space representation from a given covariant representation and taking the weak closure. From this point of view, elements in $M \rtimes_{\alpha}^{(K)} G$ can be regarded as operators of the form
\beq
	\rho^{(K)}(\mathfrak{X}) = \int_{G} \mu(g) \; \lambda^{(K)}(g) \pi_{\alpha}^{(K)}(\mathfrak{X}(g)),
\eeq
where $\mathfrak{X}: G \rightarrow M$ and $\mu(g)$ is the left invariant Haar measure on $G$. In other words, the crossed product is merely a generalization of the group algebra in which scalar coefficients are replaced by coefficients with values in $M$. The second approach is, a priori, valid only for the crossed product algebra induced by a special covariant representation we term the canonical covariant representation. Given any faithful representation $\pi: M \rightarrow B(H)$, the canonical covariant representation is formed on the Hilbert space $H_G \equiv L^2(G;H) \simeq H \otimes L^2(G)$ \cite{Takesaki2}. We denote the resulting crossed product by $M \rtimes_{\alpha} G$ and refer to it as the canonical crossed product. The canonical crossed product can be realized as a subalgebra of $M \otimes B(L^2(G))$. In fact, it is the unique invariant subalgebra of $M \otimes B(L^2(G))$ under an automorphism $\theta^{\alpha}: G \rightarrow \text{Aut}(M \otimes B(L^2(G)))$ which naturally combines the automorphism $\alpha$ with the automorphism on $B(L^2(G))$ which is induced from the right regular representation of $G$.

The latter point of view of the crossed product as an invariant algebra under a given $G$-automorphism is largely the starting point for the relationship between the crossed product and constraint quantization. Roughly speaking, it indicates that the crossed product algebra $M \rtimes_{\alpha} G$ should be interpreted as an algebraic quotient of the kinematical space $M \otimes B(L^2(G))$ under the group action $\theta^{\alpha}$. However, as we stressed, this point of view on the crossed product is strictly speaking only valid in the canonical representation. This motivates further the question of relation the crossed products $M \rtimes_{\alpha}^{(K)} G$ for different choices of covariant representation, which brings us to the top down approach to the crossed product. 

The top down approach to the crossed product was pioneered by Landstad in \cite{landstad1979duality}. His motivation was the following question: given a von Neumann algebra $A$ and a locally compact group $G$, what are the necessary and sufficient conditions for there to exist a covariant system $(M,G,\alpha)$ such that $A$ is isomorphic to the canonical crossed product $M \rtimes_{\alpha} G$? In Section \ref{sec: top down} we review Landstad's classification theorem and use it to provide an alternative point of specification for a crossed product as a triple $(A,\lambda,\delta)$ where $\lambda: \mathcal{L}(G) \rightarrow A$ is a homomorphism embedding the group von Neumann algebra inside $A$ and $\delta: A \rightarrow A \otimes \mathcal{L}(G)$ is a coaction. This perspective can largely be interpreted as a generalization of Takesaki's duality theorem for crossed products to the case of arbitrary, non-Abelian locally compact groups \cite{takesaki1973duality}. Using Landstad's theorem, we demonstrate that, for a fixed covariant system $(M,G,\alpha)$, all crossed products $M \rtimes_{\alpha}^{(K)} G$ are isomorphic to the canonical crossed product $M \rtimes_{\alpha} G$ and thus share in the constrained system interpretation for this algebra described above. 

This leads us to Section \ref{sec: cp-qrf} in which we outline the relationship between crossed product algebras, quantum reference frames and internal reference frames. We take as our definition of a quantum reference frame\footnote{The reasoning behind calling the covariant system the QRF is motivated by the top-down characterization of the crossed product. Choosing a compatible unitary representation $\lambda$ of $G$ implicates a dynamical system tailored for $\lambda$, if the underlying algebra is to be a crossed product. In light of this, our definition of the QRF is well-motivated despite not being the standard definition in the community. We note that our definition reduces to the standard one when the system algebra is fixed.} a covariant system $(M,G,\alpha)$ and demonstrate that any $G$-internal frame -- here defined as a Hilbert space $H_r$ admitting a unitary representation of the group $G$ along with a collection of covariantly transforming orientation states $\un{e}_g$ -- induces a covariant representation thereof. In other words, each QRF identifies a quotient algebra satisfying a $G$-constraint, with compatible internal frames quantifying specific measurement protocols for identify the orientation of system degrees of freedom relative to a chosen $G$-valued probe. This is formulated directly in terms of Page-Wooters reduction maps \cite{page_evolution_1983,wootters_time_1984} which implement the measurement of a chosen state's overlap with the orientations $\un{e}_g$. Since any choice of covariant representation realizes an isomorphic crossed product algebra, we conclude that the choice of internal frame has no bearing on the physics of the constrained system, only the way that it is represented. In this respect choosing an internal frame may be compared with gauge fixing -- it selects a particular representative of a gauge orbit, but any choice results in the same gauge invariant physics. 

The above observation implies that a single crossed product algebra is associated with a solitary QRF. This QRF can be represented in various ways, but the physics it describes will always be the same. To bring about a circumstance in which QRFs become truly significant we need to consider an algebra which is more complicated than a single crossed product. We provide an example of such an algebra through the simple case of a tripartite system $H = H_1 \otimes H_2 \otimes H_3$ in which each $H_i$ is an internal frame for a common group $G$. The physical operators in this case, by which we mean the operators that satisfy the $G$-constraints of the theory, can therefore be realized by splitting the kinematical space $H$ into a `system' $H_s$ and a `frame' $H_r$. Each such choice realizes a covariant system $(B(H_s), G, \text{Ad}_{U^{(s)}})$ whose associated crossed product algebra, $A_{(s \mid r)}$, has the interpretation of dressing the system operators relative to probe states built from operators in $B(H_r)$. In other words, the operators in $B(H_s)$ are made physical by conditioning their measurements on the orientations of $H_r$. The full gauge invariant subalgebra of $B(H)$ is not contained in any single crossed product algebra. Rather, we must sew together the various crossed products $A_{(s \mid r)}$ to obtain a complete accounting, with certain operators only being observable when conditioning on a particular chosen reference frame, while others are covered isomorphically in multiple frames.

Building upon the insights of the tripartite example, in Section \ref{sec: big crossed} we introduce the concept of a $G$-framed algebra. A $G$-framed algebra is an involutive Banach algebra $\mathfrak{A}$ along with a collection we term a `crossed product atlas'. In analogy with the common use of an atlas in differential geometric settings, a crossed product atlas is a collection of local `crossed product charts', 
each of which is a von Neumann subalgebra, $A_i \subset \mathfrak{A}$, isomorphic to a crossed product specified from the top down with symmetry group $H_i$. We require that the group $H_i$ is a quotient of the overall symmetry group $G$ by a subgroup $G_i$ which will be interpreted as the isotropy of the chart. The algebra $\mathfrak{A}$ is realized as a union over the crossed products contained in a given atlas up to an equivalence relation that tracks isomorphism between overlapping subalgebras of pairs of charts. In other words, $\mathfrak{A}$ is comprised of a collection of distinct QRFs along with maps that implement a change of frame whenever two frames mutually admit operators. Each QRF identifies the physical, constraint quantized physics of a particular local observer. Each operator in $\mathfrak{A}$ possesses a gauge invariant description, but nevertheless every operator may not be observable in every reference frame. In the event that $\mathfrak{A}$ can be covered by a single chart it is isomorphic to a single crossed product. But in the general case $\mathfrak{A}$ will require multiple QRFs that are not totally overlapping, and thus the $G$-framed algebra can be regarded as a natural generalization of the crossed product. Indeed, individual crossed product algebras are regarded as local quotient spaces embedded inside the larger frame-independent algebra $\mathfrak{A}$. This naturally lends itself to a discussion of the Gribov ambiguity in the context of gauge theories \cite{Langmann:1993pb,1994CMaPh.160..431M}. The $G$-framed algebra exhibits a Gribov problem in the sense that the local crossed products corresponding to different frames may not be isomorphic, leading to non-overlapping charts interpreted as incompatible gauge-fixings of the theory. 

We conclude our discussion of the $G$-framed algebra in Sections \ref{sec: de sitter} and \ref{sec: rel states}. First, we formulate the semiclassical algebra of gauge invariant observables in the static patch of de Sitter space as a $G$-framed algebra. Although this example does not take full advantage of the structure of the $G$-framed algebra, it provides a good illustration of a scenario in which having multiple QRFs is natural. This leads into Section \ref{sec: rel states} in which we briefly discuss the theory of weights, states, and entropy for $G$-framed algebras. The global structure of the algebra and the role of crossed product charts therein makes it clear that typical quantities are highly frame dependent.

\section{Crossed Product Algebras: Two Approaches} \label{sec: crossed products}
In this section we provide an introduction to crossed product algebras. We begin in Subsection \ref{sec: bottom up} by reviewing the construction of crossed products from what we term the \emph{bottom up} point of view. That is, we construct a von Neumann algebra naturally associated with a von Neumann covariant system $(M,G,\alpha)$. We make a careful observation about the possible dependence of such an algebra on a choice of covariant representation, $(K,\pi^{(K)}_{\alpha},\lambda^{(K)})$, paying special attention to the so-called canonical covariant representation induced from a faithful representation of the algebra $M$. We refer to the crossed product induced by such a representation as the {\it canonical crossed product}. In summary, we provide three different characterizations of bottom up crossed products:
\begin{enumerate}
    \item As the von Neumann algebra generated by $\pi_{\alpha}^{(K)}(M)$ and $\lambda^{(K)}(G)$,
    \item As the weak closure of an involutive Banach algebra whose elements can be interpreted as maps from the group $G$ into the von Neumann algebra $M$, and
    \item As the invariant subalgebra under a modified automorphism derived from $\alpha$.
\end{enumerate}
Strictly speaking the third characterization is unique to the canonical crossed product.

In Section \ref{sec: top down} we provide an alternative perspective of the crossed product which we term the \emph{top down} point of view. This point of view is largely inspired by the work of Landstad \cite{landstad1979duality} which provided a complete classification of crossed product algebras answering the questions, 
\begin{enumerate}
    \item Under what circumstance is a given von Neumann algebra $A$ isomorphic to a canonical crossed product algebra associated with some covariant system $(M,G,\alpha)$?
    \item If $A$ is isomorphic to such a crossed product, can we construct the associated covariant system?
\end{enumerate}
Using Landstad's classification theorem we will prove that all crossed products associated with a fixed covariant system $(M,G,\alpha)$, whether generated relative to the canonical covariant representation or not, are isomorphic to the canonical crossed product. A consequence of this observation is that all crossed products share the third characterization above as an invariant subalgebra relative to an extended $G$-automorphism. This will be of importance in Section \ref{sec: cp-qrf} when we consider the relationship between reference frames and crossed product algebras.

\subsection{Bottom-up approach to crossed product algebras} \label{sec: bottom up}

A \emph{von Neumann covariant system} is a triple $(M,G,\alpha)$ consisting of a von Neumann algebra $M$ along with a $G$-automorphism $\alpha: G \times M \rightarrow M$. A \emph{covariant representation} of the covariant system $(M,G,\alpha)$ is a triple $(K,\pi^{(K)}_{\alpha},\lambda^{(K)})$ where $K$ is a Hilbert space admitting representations $\pi^{(K)}_{\alpha}: M \rightarrow B(K)$ and $\lambda^{(K)}: G \rightarrow U(K)$ which are compatible with the automorphism $\alpha$ in the sense that
\beq \label{Covariance}
	\pi_{\alpha}^{(K)} \circ \alpha_g(x) = \text{Ad}_{\lambda^{(K)}(g)}\bigg( \pi_{\alpha}(x) \bigg). 
\eeq
In other words, the automorphism is implemented unitarily by $\lambda^{(K)}$. Given any representation $\pi: M \rightarrow B(H)$ we can construct a \emph{canonical covariant representation} $H_G \equiv L^2(G;H) \simeq H \otimes L^2(G)$ with
\begin{flalign} \label{Canonical Covariant Rep}
	&\pi_{\alpha}: M \rightarrow B(H_G), \qquad \bigg(\pi_{\alpha}(x)(\xi)\bigg)(h) \equiv \pi \circ \alpha_{h^{-1}}(x)\bigg(\xi(h)\bigg), \nonumber \\
	&\lambda: G \rightarrow U(H_G), \qquad \bigg(\lambda(h)(\xi)\bigg)(g) \equiv \xi(h^{-1}g). 
\end{flalign}
Here and henceforth, the canonical covariant representation is distinguished by the absence of a superscript. 

Among the original motivations behind the crossed product was to associate a von Neumann algebra with a given covariant system \cite{takesaki1973duality}. Given the canonical covariant representation, we define the \emph{canonical crossed product algebra} $M \rtimes_{\alpha} G$ as the von Neumann algebra generated by $\pi_{\alpha}(x)$ and $\lambda(g)$ in the weak operator topology induced by $H_G$. More generally, one can construct a crossed product starting with any covariant representation. Let $(K, \pi^{(K)}, \lambda^{(K)})$ be a covariant representation of the covariant system $(M,G,\alpha)$. Then, we define the $K$-\emph{crossed product algebra} by\footnote{Here, $\vee$ is the von Neumann algebraic union.}
\beq \label{K crossed product}
    M \rtimes^{(K)}_{\alpha} G \equiv \pi^{(K)}(M) \vee \lambda^{(K)}(G).
\eeq    
As it is defined, it is not clear how $M \rtimes_{\alpha}^{(K)} G$ and $M \rtimes_{\alpha}^{(K')} G$ are related for distinct representations $K$ and $K'$. In Section \ref{sec: top down} we will address this point directly. In Section \ref{sec: cp-qrf} we will demonstrate how standard quantum reference frames can be interpreted as generating distinct covariant representations given a fixed covariant system. 


There exists an alternative definition of the crossed product introduced by Haagerup \cite{haagerup1978dualI,haagerup1978dualII}. Firstly, let
\beq \label{MG}
	M_G \equiv \{\mathfrak{X}: G \rightarrow M \; | \; \mathfrak{X} \text{ is strongly continuous, and has compact support}\}. 
\eeq
The set $M_G$ can be turned into an involutive Banach algebra by endowing it with a product and involution as given by\footnote{Here $\delta(g)$ is the module function which tracks the non right-invariance of the left-invariant Haar measure $\mu$.}
\begin{flalign} \label{Structures on MG}
	&\bigg(\mathfrak{X} \mathfrak{Y}\bigg)(g) \equiv \int_{G} \mu(h) \;\alpha_{h}\bigg(\mathfrak{X}(gh)\bigg)\mathfrak{Y}(h^{-1}), \nonumber \\
	&\bigg(\mathfrak{X}^*\bigg)(g) \equiv \delta(g^{-1}) \alpha_{g^{-1}}\bigg(\mathfrak{X}(g^{-1})\bigg)^*. 
\end{flalign}
Given \emph{any} covariant representation $(K,\pi_{\alpha}^{(K)}, \lambda^{(K)})$ we can construct a representation $\rho^{(K)}: M_G \rightarrow B(K)$ as given by
\beq \label{Haagerup rep}
	\rho^{(K)}(\mathfrak{X}) \equiv \int_G \mu(g) \; \lambda^{(K)}(g) \pi^{(K)}_{\alpha}\bigg(\mathfrak{X}(g)\bigg).
\eeq
Haagerup demonstrated that the image algebra $\rho^{(K)}(M_G)$ is dense in the von Neumann algebra generated by $\pi^{(K)}_{\alpha}(M)$ and $\lambda^{(K)}(G)$ (in the weak operator topology induced by $K$). Thus, in general $\rho^{(K)}(M_G)'' = M \rtimes_{\alpha}^{(K)} G$.\footnote{Endowing the algebra $M_G$ with a compatible ultraweak topology one can realize the so-called $C^*$ crossed product of $M$ by $G$ \cite{doplicher1966covariance}. This algebra was originally introduced to study the space of covariant representations of a given covariant system $(M,G,\alpha)$.}

Finally, restricting our attention to the canonical crossed product algebra $M \rtimes_{\alpha} G$ we can describe yet a third characterization of the crossed product which is valuable for understanding its role in implementing constraints. Firstly, let us note that $M \rtimes_{\alpha} G$ can be realized as a subalgebra of $M \otimes B(L^2(G))$ \cite{van1978continuous}. On the space $M \otimes B(L^2(G))$ we define an automorphism
\beq \label{extended action}
	\theta^{\alpha} \equiv \alpha \otimes \text{Ad}_{r}: G \rightarrow \text{Aut}\bigg(M \otimes B(L^2(G))\bigg),
\eeq
where here,
\beq
	r: G \rightarrow U(L^2(G)), \qquad \bigg(r(g)(\xi)\bigg)(h) \equiv \delta(g)^{1/2} \xi(hg)
\eeq
is the regular right action on $L^2(G)$. 

Alternatively, we may denote by $\rho: G \rightarrow U(H_G)$ the right representation lifted to the Hilbert space $H_G$:
\beq \label{Right rep on HG}
	\bigg(\rho(g) \xi\bigg)(h) \equiv \delta(g)^{1/2} \xi(hg). 
\eeq
Similarly, let us also denote by $\tilde{\alpha} \equiv \alpha \otimes \mathbb{1}: G \rightarrow \text{Aut}(M \otimes B(L^2(G)))$ the extension of the automorphism $\alpha$ now acting on $M \otimes B(L^2(G))$. In terms of these definitions the automorphism \eqref{extended action} can equivalently be written as
\beq
    \theta^{\alpha} = \tilde{\alpha} \circ \text{Ad}_{\rho} = \alpha \otimes \text{Ad}_r. 
\eeq

It can be shown that the crossed product algebra $M \rtimes_{\alpha} G$ is the unique invariant subalgebra of $M \otimes B(L^2(G))$ under the automorphism $\theta^{\alpha}$:
\beq \label{Invariance of CP algebra}
	M \rtimes_{\alpha} G = \bigg(M \otimes B(L^2(G))\bigg)^{\theta^{\alpha}} \equiv \{\mathcal{O} \in M \otimes B(L^2(G)) \; | \; \theta^{\alpha}_g(\mathcal{O}) = \mathcal{O}, \; \forall g \in G\}. 
\eeq
The invariance of $\lambda(g)$ under \eqref{extended action} is trivially obtained from the commutation of the left and right actions on $G$. Thus, to demonstrate that $M \rtimes_{\alpha} G \subset \bigg(M \otimes B(L^2(G;\mu))\bigg)^{\theta^{\alpha}}$ it remains only to show that $\theta^{\alpha}_g(\pi_{\alpha}(x)) = \pi_{\alpha}(x)$ for every $x \in M, g \in G$. 

To make this observation, let us first compute explicitly $\text{Ad}_{\rho(h)}(\pi_{\alpha}(x)) = \rho(h) \pi_{\alpha}(x) \rho(h^{-1})$. To do so we will act with this operator on a generic element $\xi \in H_G$ in a sequence of two steps. Firstly we have
\begin{flalign} \label{Right action of right action on pi alpha}
	\bigg(\pi_{\alpha}(x)\rho(h^{-1}) \xi\bigg)(g) &= \pi_{\alpha}(x) \bigg(\rho(h^{-1}) \xi\bigg)(g) \nonumber \\
	&= \delta(h)^{-1/2} \pi \circ \alpha_{g^{-1}}(x) \bigg(\xi(gh^{-1})\bigg). 
\end{flalign} 
Here we have used \eqref{Right rep on HG} and \eqref{Canonical Covariant Rep}. Next we can compute
\begin{flalign} \label{Right adjoint action on pi alpha}
	\bigg(\rho(h) \pi_{\alpha}(x) \rho(h^{-1}) \xi\bigg)(g) &= \rho(h) \bigg(\pi_{\alpha}(x) \rho(h^{-1})\xi \bigg)(g) \nonumber \\
	&= \delta(h)^{1/2} \bigg(\pi_{\alpha}(x) \rho(h^{-1}) \xi\bigg)(gh) \nonumber \\
	&= \delta(h)^{1/2} \delta(h)^{-1/2} \pi \circ \alpha_{h^{-1} g^{-1}}(x) \bigg(\xi((gh) h^{-1})\bigg) \nonumber \\
	&= \pi \circ \alpha_{h^{-1}} \circ \alpha_{g^{-1}}(x) \bigg(\xi(g)\bigg). 
\end{flalign}
At the same time, supposing that $\alpha$ is unitarily implemented\footnote{That is
\beq
    \pi \circ \alpha_g(x) = V_g \pi(x) V_g^{\dagger}.
\eeq
Note that this makes no assumption as to whether $V(G) \subset \pi(M)$, e.g. the automorphism $\alpha$ needn't be inner. We will revisit this point in Section \ref{sec: top down}.
} on $H$ via the representation $V: G \rightarrow U(H)$ we can write
\beq
	\tilde{\alpha}_g(\mathcal{O}) \equiv \text{Ad}_{V_g \otimes \mathbb{1}}(\mathcal{O}), \qquad \mathcal{O} \in M \otimes B(L^2(G)). 
\eeq
Acting on $\pi_{\alpha}(x) \in M \rtimes_{\alpha} G \subset M \otimes B(L^2(G))$ we have
\begin{flalign} \label{Extended alpha on pi alpha}
	\bigg(\tilde{\alpha}_h(\pi_{\alpha}(x)) \xi \bigg)(g) &= \bigg((V_h \otimes \mathbb{1}) \pi_{\alpha}(x) (V_{h^{-1}} \otimes \mathbb{1}) \xi \bigg)(g) \nonumber \\
	&= V_h \pi \circ \alpha_{g^{-1}}(x) V_{h^{-1}} \bigg(\xi(g)\bigg) = \pi \circ \alpha_h \circ \alpha_{g^{-1}}(x) \bigg(\xi(g)\bigg). 
\end{flalign}
Here we have used the fact that $V_h \pi(x) V_{h^{-1}} = \pi \circ \alpha_h(x)$. Comparing \eqref{Right adjoint action on pi alpha} with \eqref{Extended alpha on pi alpha} we see that 
\beq
	\text{Ad}_{\rho(h)}(\pi_{\alpha}(x)) = \tilde{\alpha}_{h^{-1}}(\pi_{\alpha}(x)),
\eeq
which further implies that
\beq \label{Final invariance of pi alpha}
	\tilde{\alpha}_{h} \circ \text{Ad}_{\rho(h)}(\pi_{\alpha}(x)) = \pi_{\alpha}(x),
\eeq
as desired. 

This completes one half of the proof of the commutation theorem, namely demonstrating that every element in $M \rtimes_{\alpha} G$ is invariant under the automorphism $\theta^{\alpha}$. The second half of the proof is to show that every invariant element of $M \otimes B(L^2(G))$ is of this form. This exercise is less informative and so we refer to \cite{van1978continuous} for a proof.\footnote{In Appendix \ref{app: eps} an analogous computation is carried out in the context of the extended phase space, which (in a local trivialization) may be interpreted as the classical analog of the crossed product. The computation in \eqref{Invariance of Dressed Observables} is remarkably similar in form to the commutation theorem. Essentially, one shows that dressed observables in the extended phase space are invariant under the right action of the structure group $G$ because they are acted upon by compensating actions analogous to $\tilde{\alpha}$ and $\text{Ad}_{\rho}$.} 

\subsection{Top-down approach to crossed product algebras} \label{sec: top down}

In the previous subsection we have stressed the importance of crossed product algebras for studying systems with explicit constraints. Up to this point our perspective has been `bottom up' in the sense that we have started with an explicit covariant system $(M,G,\alpha)$ and subsequently formed the associated crossed product algebra $M \rtimes_{\alpha} G$. In this section we will introduce a `top-down' approach in which we begin with an arbitrary von Neumann algebra $A$ and a fixed locally compact group $G$ and determine the necessary and sufficient conditions under which $A$ is isomorphic to the crossed product algebra of some covariant system $(M,G,\alpha)$. In addition, provided $A$ is isomorphic to some crossed product, we present a recipe for identifying a covariant system $(M,G,\alpha)$ for which $A \simeq M \rtimes_{\alpha} G$. 

To begin, let us highlight some properties of crossed product algebras which we have not stressed in the previous subsection. Let $(M,G,\alpha)$ be a covariant system, and $(H_G, \pi_{\alpha}, \lambda)$ the canonical covariant representation induced from a faithful representation $\pi: M \rightarrow B(H)$. Unless otherwise specified, the following analysis is specialized to the canonical crossed product algebra. We first make the following observation: if $\alpha$ is strictly inner, then $M \rtimes_{\alpha} G \simeq M \otimes \mathcal{L}(G)$.\footnote{Here, $\mathcal{L}(G)$ is the group von Neumann algebra associated with $G$ as reviewed in Appendix \ref{app: group vN}.} Because $\alpha$ is inner, there exists a homomorphism\footnote{Here $U(M) \subset M$ is the set of unitary elements in $M$.} $u: G \rightarrow U(M)$ such that $\alpha_g(x) = u_g x u_g^*$. Under this state of affairs we can define a unitary element $U \in U(H_G)$ such that
\beq \label{U map}
	(U\xi)(g) \equiv \pi(u_g)(\xi(g)), \;\; \forall \xi \in H_G. 
\eeq
Then the covariant representations $\pi_{\alpha}$ and $\lambda$ can be expressed
\beq \label{Covariant Rep as tensor product}
	\pi_{\alpha}(x) = U^{\dagger} (\pi(x) \otimes \mathbb{1}) U, \qquad \lambda(g) = U^{\dagger} (\pi(u_g) \otimes \ell(g)) U. 
\eeq
The crossed product is given by $\pi_{\alpha}(M) \vee \lambda(G)$ and thus is unitarily equivalent to the von Neumann algebra generated by $\pi(x) \otimes \mathbb{1}$ and $\pi(u_g) \otimes \ell(g)$. 

More generally, so long as $\alpha$ is implemented unitarily on $H$ we can still write formulae which are reminiscent of \eqref{U map} and \eqref{Covariant Rep as tensor product}. That is, we suppose that there exists a unitary representation\footnote{We recall that any automorphism will automatically be unitarily implemented in a standard representation of a given von Neumann algebra \cite{Takesaki2}.} $V: G \rightarrow U(H)$ such that
\beq
	\pi \circ \alpha_g(x) = V_g \pi(x) V_g^{\dagger}.
\eeq 
Then, we can redefine
\beq
	(U\xi)(g) \equiv V_g(\xi(g)),
\eeq
and by extension write
\beq \label{Canonical Covariant Rep in tensor product form}
	\pi_{\alpha}(x) = U^{\dagger}(\pi(x) \otimes \mathbb{1}) U, \qquad \lambda(g) = U^{\dagger}(V_g \otimes \ell(g)) U. 
\eeq
The caveat of course is that $V_g$ needn't be in $\pi(M)$ for each $g$ and hence $V_g \otimes \ell(g)$ needn't be in $\pi(M) \otimes \lambda(G)$. 

One upshot of the previous discussion is that to realize non-factorizable crossed product algebras we need to consider outer actions. The second observation, however, is that we can use the inner automorphism $\text{Ad}_{\lambda(g)}$ to reframe the analysis of $M \rtimes_{\alpha} G$ in terms of a tensor product algebra. In particular, we concentrate our analysis on the Hilbert space $L^2(G,H_G) \simeq L^2(G \times G,H) \simeq H_G \otimes L^2(G)$. On this space we can always define a unitary operator $W \in U(L^2(G,H_G))$ by
\beq \label{W operator}
	(W \xi)(g,h) = \xi(g,gh), \qquad \xi \in L^2(G,H_G). 
\eeq
The operator $W$ implements a mapping $\delta: M \rtimes_{\alpha} G \rightarrow (M \rtimes_{\alpha} G) \otimes \mathcal{L}(G)$ given by
\beq \label{Coaction}
	\delta(\mathfrak{X}) \equiv W^{\dagger} \bigg(\mathfrak{X} \otimes \mathbb{1}\bigg) W.
\eeq
The map defined in \eqref{Coaction} is a normal isomorphism and is referred to as a coaction on $M \rtimes_{\alpha} G$.\footnote{The map $\delta$ is called a coaction because it induces a representation of the predual $\mathcal{L}(G)_*$ on the predual $(M \rtimes_{\alpha} G)_*$.} By analogy to the previous argument $(M \rtimes_{\alpha} G) \otimes \mathcal{L}(G)$ can be interpreted as the `double' crossed product $(M \rtimes_{\alpha} G) \rtimes_{\text{Ad}_{\lambda}} G$.\footnote{In the case that $G$ is an Abelian group the coaction $\delta$ is equivalent to an action by the Pontryagin dual group. The `double' crossed product is then explicitly the double crossed product of $M$ first by the group action and then by the dual group action.}

It is instructive to evaluate \eqref{Coaction} when acting on the generators of the crossed product. By a straightforward computation one can show,
\beq \label{Coaction on M and G}
	\delta(\pi_{\alpha}(x)) = \pi_{\alpha}(x) \otimes \mathbb{1}, \qquad \delta(\lambda(g)) = \lambda(g) \otimes \ell(g). 
\eeq
In words, the image of $M$ inside the crossed product is `invariant' under the coaction, while the image of $G$ obtains a new tensor factor corresponding to the regular left action. In the specific case that $M = \mathbb{C}$ so that the crossed product is merely isomorphic to $\mathcal{L}(G)$ the coaction defines a map $\delta_G: \mathcal{L}(G) \rightarrow \mathcal{L}(G)^{\otimes 2}$ by
\beq \label{G coaction}
	\delta_G(\ell(g)) = \ell(g) \otimes \ell(g). 
\eeq
From eq. \eqref{Coaction on M and G} we find\footnote{Here $i$ denotes the appropriate identity map relative to the tensor factor it acts upon.}
\begin{flalign}
	(\delta \otimes i) \circ \delta(\pi_{\alpha}(x)) &= \pi_{\alpha}(x) \otimes \mathbb{1} \otimes \mathbb{1} = (i \otimes \delta_G) \circ \delta(\pi_{\alpha}(x)) \nonumber \\
	(\delta \otimes i) \circ \delta(\lambda(g)) &= \lambda(g) \otimes \ell(g) \otimes \ell(g) = (i \otimes \delta_G) \circ \delta(\lambda(g)). 
\end{flalign}
Since $\pi_{\alpha}(M)$ and $\lambda(G)$ together generate the full crossed product algebra we conclude that
\beq \label{Coaction relation}
	(\delta \otimes i) \circ \delta = (i \otimes \delta_G) \circ \delta. 
\eeq

We are now prepared to state Landstad's classification of the space of $G$-crossed product algebras \cite{landstad1979duality}: 
\begin{theorem}[Landstad's Classification Theorem]
Let $A$ be a von Neumann algebra and $G$ a locally compact group. The algebra $A$ is isomorphic to a (regular) crossed product $M \rtimes_{\alpha} G$ for some covariant system $(M,G,\alpha)$ if and only if there exists a continuous homomorphism $\lambda: G \rightarrow A$ and a coaction $\delta: A \rightarrow A \otimes \mathcal{L}(G)$ satisfying
\begin{flalign}
	\delta(\lambda(g)) &= \lambda(g) \otimes \ell(g), \; \forall g \in G, \nonumber \\
	(\delta \otimes i) \circ \delta(\mathfrak{X}) &= (i \otimes \delta_G) \circ \delta(\mathfrak{X}), \; \forall \mathfrak{X} \in A. 
\end{flalign}
What's more, the covariant system $(M,G,\alpha)$ is determined uniquely by $\delta$ and $\lambda$ as
\beq
	M = \{\mathfrak{X} \in A \; | \; \delta(\mathfrak{X}) = \mathfrak{X} \otimes \mathbb{1}\}, \qquad \alpha: G \rightarrow \text{Aut}(M), \; \alpha_g(\mathfrak{X}) = \lambda_g \mathfrak{X} \lambda_g^*, \; g \in G, \mathfrak{X} \in M. 
\eeq
\end{theorem}
Landstad's theorem establishes that a crossed product algebra can be specified uniquely by the triple $(A,\delta,\lambda)$. We refer to this as the \emph{top-down} specification of a crossed product, in contrast to the covariant system $(M,G,\alpha)$ which we refer to as a \emph{bottom-up} specification.

Let $(A,\lambda,\delta)$ be a crossed product algebra specified from the top down with associated covariant system $(M,G,\alpha)$. As we have reviewed in App. \ref{app: group vN}, the algebra $\mathcal{L}(G)$ comes equipped with a natural faithful, semi-finite, normal weight $\gamma$ called the Plancherel weight. On the algebra $A \otimes \mathcal{L}(G)$, the Plancherel weight defines a slice map $P_{\gamma}: A \otimes \mathcal{L}(G) \rightarrow A$ such that
\beq
	\varphi \circ P_{\gamma}(\mathcal{X}) = \varphi \otimes \gamma(\mathcal{X}), \; \mathcal{X} \in A \otimes \mathcal{L}(G), \varphi \in A_*. 
\eeq
In other words, $P_{\gamma}$ can be interpreted as a partial trace on $A \otimes \mathcal{L}(G)$ with respect to the Plancherel weight. Composing the map $P_{\gamma}$ with the map $\delta$ we obtain a map $T_{\gamma}: P_{\gamma} \circ \delta: A \rightarrow A$. The map $T_{\gamma}$ is closely related to Haagerup's operator-valued weight introduced in \cite{haagerup1978dualI}. In particular, $\text{im}(T_{\gamma}) = M$ and
\beq
	T_{\gamma}(x \mathfrak{X} y) = x T_{\gamma}(\mathfrak{X}) y, \qquad x,y \in M, \mathfrak{X} \in A. 
\eeq

The map $T_{\gamma}$ provides an alternative point of view on the top-down specification of the crossed product in terms of the sequence of maps:
\begin{equation} \label{Sequence}
\begin{tikzcd}
\mathcal{L}(G)
\arrow{r}{\lambda} 
& 
A
\arrow{r}{T} 
& 
M.
\end{tikzcd}
\end{equation}
In other words, we may regard the top down specification of the crossed product as a triple $(A,\lambda,T_{\gamma})$, where $\lambda$ is a homomorphism embedding the group von Neumann algebra into $A$ and $T_{\gamma}$ is an operator-valued weight projecting $A$ into the von Neumann algebra $M$. 

We note in passing that this perspective of the top down specification of the crossed product is very evocative of the relationship between the crossed product and the extended phase space \cite{Klinger:2023tgi}. The sequence \eqref{Sequence} is analogous to that which defines the Atiyah Lie algebroid associated with the extended phase space:
\begin{equation} \label{Short Exact Sequence ALA}
\begin{tikzcd}
0
\arrow{r} 
& 
L
\arrow{r}{j} 
\arrow[bend left]{l} 
& 
A
\arrow{r}{\rho} 
\arrow[bend left]{l}{\omega}
& 
TX
\arrow{r} 
\arrow[bend left]{l}{\sigma}
&
0\,.
\arrow[bend left]{l} 
\end{tikzcd}
\end{equation}
Here, in the case of the Lie algebroid, we have specified a second sequence in terms of the maps $(\sigma,\omega)$ which together encode the data of a principal connection. In Section \ref{sec: QRFs} we will specify a lower sequence on the diagram \eqref{Sequence} with the interpretation of specifying a trivial splitting, i.e., the algebraic analog of a flat connection. In future work we intend to explore the interpretation of non-trivial splittings of the sequence \eqref{Sequence} in an effort to interpret holonomy and curvature in the context of crossed product algebras as we will  discuss in Sec.~\ref{sec: curv} below. 

Given Landstad's theorem we can now revisit the question posed in Section \ref{sec: bottom up} about the relationship between crossed product algebras derived from non-canonical covariant representations. Let $(M,G,\alpha)$ be a fixed covariant system and $(K,\pi_{\alpha}^{(K)}, \lambda^{(K)})$ a covariant representation. The $K$-crossed product algebra, $M \rtimes_{\alpha}^{(K)} G$, is the von Neumann algebra generated by $\pi_{\alpha}^{(K)}(M)$ and $\lambda^{(K)}(G)$. We may now apply Landstad's theorem directly to the algebra $M \rtimes_{\alpha}^{(K)} G$. Clearly this algebra admits a homomorphism $\lambda^{(K)}: G \rightarrow M \rtimes_{\alpha}^{(K)} G$. Moreover, we can construct a coaction $\delta^{(K)}: M \rtimes_{\alpha}^{(K)} G \rightarrow M \rtimes_{\alpha}^{(K)} G \otimes \mathcal{L}(G)$ by specifying its action on the generators as
\beq
    \delta^{(K)}(\pi_{\alpha}^{(K)}(x)) \equiv \pi_{\alpha}^{(K)}(x) \otimes \mathbb{1}, \qquad \delta^{(K)}(\lambda^{(K)}(g)) \equiv \lambda^{(K)}(g) \otimes \ell(g). 
\eeq
By analogy to \eqref{Coaction on M and G} it is easy to conclude that this coaction satisfies the conditions of Landstad's theorem and thus we conclude that $M \rtimes_{\alpha} G$ is isomorphic to some canonical crossed product. The covariant system associated with this crossed product is $(\pi_{\alpha}^{(K)}(M), G, \text{Ad}_{\lambda^{(K)}})$, however as $\pi_{\alpha}^{(K)}(M)$ is isomorphic to $M$ and $\lambda^{(K)}$ implements the automorphism $\alpha$ on $\pi_{\alpha}^{(K)}(M)$ the aforementioned covariant system is equivalent to $(M,G,\alpha)$. Thus, we conclude that
\beq \label{All K crossed products are isomorphic}
    M \rtimes^{(K)}_{\alpha} G \simeq \pi^{(K)}(M) \rtimes_{\text{Ad}_{\lambda^{(K)}}} G \simeq M \rtimes_{\alpha} G.
\eeq

Eqn. \eqref{All K crossed products are isomorphic} implies that all $K$-crossed products associated with a given covariant system $(M,G,\alpha)$ are isomorphic to the canonical crossed product $M \rtimes_{\alpha} G$, and thus are also isomorphic to each other. In particular, this means that every $K$-crossed product shares in the invariant operator interpretation of the crossed product associated with the commutation theorem.  

\section{Algebraic Quantum Reference Frames} \label{sec: QRFs}

In Section \ref{sec: crossed products} we explored several different interpretations for crossed product algebras. In Section \ref{sec: bottom up} we raised the question of whether crossed product algebras generated from different covariant representations of a common covariant system are equivalent. Using Landstad's classification theorem, we answered this question in the affirmative at the end of Section \ref{sec: top down}. In this section we will provide an interpretation for distinct covariant systems as quantum reference frames (QRFs) with non-canonical covariant representations of a given covariant system corresponding to an internal frame. From this point of view, the statement that all $K$-crossed product algebras are isomorphic implies that the crossed product algebras induced from different internal frames are always equivalent, provided these reference frames are appended to common `system' algebras, i.e., common QRFs. This can be interpreted as a statement of gauge-invariance. 

The aforementioned conclusion, that all internal frames give rise to the same crossed product algebras when appended to a common QRF, inspires the observation that the relevance of QRFs can only be fully appreciated in the context of an algebra which is more sophisticated than a single crossed product. This observation is substantiated by the example of a tripartite quantum system in which the set of physical operators is not covered by a single crossed product algebra, but rather by a collection of distinct crossed product algebras along with transition maps identifying these algebras wherever they intersect. This inspires the introduction of a new algebraic object we refer to as a \emph{$G$-framed algebra} which one may interpret as a kind of non-commutative analog of an orbifold\footnote{Orbifolds are technically quotients of manifolds under a finite group action, but in this work we consider general locally compact groups $G$.} with fundamental symmetry group $G$. We provide a discussion of the utility of the $G$-framed bundle in articulating the Gribov problem in a fully quantum context in Subsection~\ref{sec: big crossed}, and in delineating between frame-dependent and frame-independent quantities in Subsection~\ref{sec: rel states}.

\subsection{Crossed products and quantum reference frames} \label{sec: cp-qrf}

In this subsection we demonstrate that, given a fixed quantum reference frame, any internal frame can be used to induce a covariant representation and by extension a crossed product algebra. By eq. \eqref{All K crossed products are isomorphic} this crossed product will be isomorphic to a canonical crossed product and thus the internal frame may be interpreted as dressing operators to implement a particular constraint. In this context we can construct this constraint very explicitly. 

In the following, we shall take as our definition of a QRF as the specification of a fixed covariant system $(M,G,\alpha)$; in the following section the reasoning behind this definition will become clear. Relative to such a choice, we define an internal frame as follows:
\begin{definition}[Internal Reference Frame]
	Given a locally compact group $G$, a $G$-internal frame (or simply internal frame for short) is a Hilbert space $H_r$ which admits a unitary representation $U^{(r)}: G \rightarrow U(H_r)$ along with a (potentially overcomplete) basis of \emph{orientation states} $\{\un{e}^{(r)}_g \in H_r \}_{g \in G}$ transforming covariantly under $U^{(r)}$,
	\beq \label{Covariance of Orientation States}
		U^{(r)}(g) \un{e}^{(r)}_{h} = \un{e}^{(r)}_{gh}. 
	\eeq
Although it is not strictly necessary, we suppose that the states $\un{e}^{(r)}_g$ are delta-function normalized so that
\beq
	g_r\bigg(\un{e}^{(r)}_g, \un{e}^{(r)}_h\bigg) = \delta(g, h),
\eeq
with $\delta(g,h)$ the delta function relative to the left-invariant Haar measure on $G$. 
\end{definition}

Now, suppose that $(M,G,\alpha)$ is a von Neumann covariant system, that is a QRF, and that $M$ admits a faithful representation $\pi: M \rightarrow B(H_s)$ on a `system' Hilbert space $H_s$. Then, we have the following claim:
\begin{theorem}[Covariant Representation Induced by an Internal Frame]
	For any internal frame $H_r$ the Hilbert space $H \equiv H_s \otimes H_r$ admits a covariant representation of the covariant system $(M,G,\alpha)$. 
\end{theorem}
To prove this theorem let us make two observations. First, by assumption of (over)completeness, the orientation states define a resolution of the identity:
\beq
	\mathbb{1}_{H_r} = \int_{G} \mu(g) \; g_r(\un{e}^{(r)}_g, \cdot) \; \un{e}^{(r)}_g. 
\eeq
Thus, any vector $\Psi \in H$ can be decomposed as
\beq \label{Resolution of states via QRF}
	\Psi = (\mathbb{1}_{H_s} \otimes \mathbb{1}_{H_r}) \Psi = \int_G \mu(g) \; g_r(\un{e}^{(r)}_g, \Psi) \otimes \un{e}^{(r)}_g \equiv \int_{G} \mu(g) \; f^{(s \mid r)}_{\Psi}(g) \otimes \un{e}^{(r)}_g. 
\eeq
The mapping $R: G \times H \rightarrow H_s$ which identifies the coefficients of the expansion in \eqref{Resolution of states via QRF} is called the Page-Wooters map~\cite{page_evolution_1983,wootters_time_1984, Page:1989br, hoehn_trinity_2021}:
\beq
	R_g(\Psi) = g_r(\un{e}^{(r)}_g, \Psi) \equiv f^{(s \mid r)}_{\Psi}(g),
\eeq
and can be interpreted as measuring the state $\Psi$ along the projection induced by $\un{e}^{(r)}_g$. Notice that \eqref{Resolution of states via QRF} identifies elements $\Psi \in H$ with square integrable maps from $G$ into $H_s$, $f^{(s \mid r)}_{\Psi} \in L^2(G,H_s)$. In this respect, the fact that $H_r$ admits a basis of orientation states labeled by group elements implies that $H$ is (at least) a Hilbert subspace of the canonical covariant representation space $H \subset L^2(G,H_s)$.  

The second observation is that the unitary representation $U^{(r)}$ induces a unitary representation $\lambda^{(r)} \equiv \mathbb{1}_{H_s} \otimes U^{(r)}: G \rightarrow U(H)$. Moreover, \eqref{Covariance of Orientation States} implies that 
\begin{flalign} \label{Frame induced unitary}
	\lambda^{(r)}(h) \Psi &= \int_{G} \mu(g) \; f^{(s \mid r)}_{\Psi}(g) \otimes U^{(r)}(h) \un{e}^{(r)}_g \nonumber \\
	&= \int_{G} \mu(g) \; f^{(s \mid r)}_{\Psi}(g) \otimes \un{e}^{(r)}_{hg} \nonumber \\
	&= \int_{G} \mu(k) \; f^{(s \mid r)}(h^{-1} k) \otimes \un{e}^{(r)}_k. 
\end{flalign}
To move from the second to the third line in \eqref{Frame induced unitary} we have used the left invariance of the Haar measure. Thus, we conclude that the unitary representation $\lambda^{(r)}$ acts on $H$ in  precisely the same way as the left regular representation of $L^2(G,H_s)$ provided we restrict our attention to the coefficients $f^{(s \mid r)}_{\Psi}$. 

Having made these observations let us now propose $\pi^{(r)}_{\alpha}: M \rightarrow B(H)$ such that
\beq \label{Frame dressing}
	\pi_{\alpha}^{(r)}(x) \Psi \equiv \int_{G} \mu(g) \; \pi \circ \alpha_{g^{-1}}(x)\bigg(f^{(s \mid r)}_{\Psi}(g)\bigg) \otimes \un{e}^{(r)}_g. 
\eeq
Of course, $\pi_{\alpha}^{(r)}$ is modeled on the canonical representation \eqref{Canonical Covariant Rep}. It is therefore straightforward to show that
\beq
	\pi_{\alpha}^{(r)} \circ \alpha_g(x) = \text{Ad}_{\lambda^{(r)}(g)}\bigg(\pi_{\alpha}^{(r)}(x)\bigg).
\eeq 
Thus, we conclude that $(H, \pi_{\alpha}^{(r)}, \lambda^{(r)})$ is a covariant representation of the covariant system $(M,G,\alpha)$, completing the proof of our theorem.

Recalling the general construction under eqn. \eqref{Canonical Covariant Rep}, we can use the covariant representation $(H,\pi_{\alpha}^{(r)},\lambda^{(r)})$ to construct a (non-canonical) crossed product:
\beq
	M \rtimes_{\alpha}^{(r)} G \equiv \pi_{\alpha}^{(r)}(M) \vee \lambda^{(r)}(G). 
\eeq
Although \eqref{All K crossed products are isomorphic} implies that this crossed product is isomorphic to the canonical crossed product, we will find it advantageous to continue to stress the representation $H_r$. In later discussions, we will interpret an internal frame as a form of gauge fixing, and so one may interpret the internal frame label as identifying such a choice. 

By an argument completely analogous to eqn. \eqref{extended action} -- \eqref{Final invariance of pi alpha} it can be shown that all of the elements of $M \rtimes_{\alpha}^{(r)} G$ are invariant under a modified automorphism induced from $\alpha$ and $\un{e}^{(r)}_g$. To be precise, first let us define the right representation $\overline{U^{(r)}}: G \rightarrow U(H_r)$ by its action on the orientation states, i.e., such that
\beq
	\overline{U^{(r)}}(g) \un{e}^{(r)}_h \equiv \delta(h)^{-1/2} \un{e}^{(r)}_{hg^{-1}}.
\eeq
Then, we can define a right representation $\rho^{(r)} \equiv \mathbb{1} \otimes \overline{U^{(r)}}: G \rightarrow U(H)$ which acts as
\begin{flalign}
	\rho^{(r)}(h) \Psi &= \int_{G} \mu(g) \; f^{(s \mid r)}_{\Psi}(g) \otimes \overline{U^{(r)}}(h) \un{e}^{(r)}_{g} \nonumber \\
	&= \int_{G} \mu(g) \; \delta(h)^{-1/2} f^{(s \mid r)}_{\Psi}(g) \otimes  \un{e}^{(r)}_{gh^{-1}} \nonumber \\
	&= \int_{G} \mu(k) \; \delta(h)^{1/2} f^{(s \mid r)}_{\Psi}(kh) \otimes \un{e}^{(r)}_k. 
\end{flalign}
Here, we have used the fact that $\mu(kh) = \delta(h) \mu(k)$. Thus we conclude that the induced right representation $\rho^{(r)}$ reproduces \eqref{Right rep on HG} when acting on the components $f^{(s \mid r)}_{\Psi}(g)$. 

Denoting by $\tilde{\alpha} \equiv \alpha \otimes \mathbb{1}$ the extension of the automorphism $\alpha$ to the algebra $M \otimes B(H_r)$ we may therefore define an extended automorphism action
\beq
	\theta^{(r)} \equiv \tilde{\alpha} \circ \text{Ad}_{\rho^{(r)}}: G \rightarrow \text{Aut}(M \otimes B(H_r)).
\eeq
Then, by the argument laid out between \eqref{extended action} -- \eqref{Final invariance of pi alpha} we conclude that
\beq
	\theta^{(r)}(\pi_{\alpha}^{(r)}(x)) = \pi_{\alpha}^{(r)}(x), \qquad \theta^{(r)}(\lambda^{(r)}(g)) = \lambda^{(r)}(g), \; \forall x \in M, g \in G,
\eeq
which implies that $M \rtimes_{\alpha}^{(r)} G$ is contained in the fixed point subalgebra of $M \otimes B(H_r)$ under the extended action $\theta^{(r)}$. Thus, we have shown that the relational crossed products $M \rtimes_{\alpha}^{(r)} G$ also retain the constraint based interpretation of the canonical crossed product. 

At the end of the day, Landstad's theorem provides a justification for the aforementioned observations. Any \emph{fixed} covariant system $(M,G,\alpha)$ with internal frame $H_r$ is isomorphic to the canonical crossed product induced by that covariant system. Nevertheless, as we will now demonstrate, a single kinematical algebra may possess many different non-isomorphic covariant systems, i.e., many different QRFs, which are only compatible with select internal frames. The ability to pass from a fixed covariant system equipped with a chosen internal frame to a crossed product algebra will therefore provide us with a rigorous tool for comparing and mapping between the physics of different QRFs. This is crucial if we want to fill out the complete algebra of physical observables in a given system.

\subsection{Systems with multiple reference frames} \label{sec: tripartite example}

In the previous subsection we presented an approach to realizing crossed product algebras directly from given covariant systems with compatible internal frames. In this subsection we introduce a natural example in which a given kinematical system possesses many possible covariant systems. In this case one has a choice as to how they wish to divide degrees of freedom into system and probe. Choosing different QRFs identify different physical operators. In this respect, the physical operators identified by any individual QRF may not be sufficient to cover the full set of physical operators that exist inside of the original kinematical algebra. To counter this observation, we will introduce an approach to sewing together multiple QRFs and their associated crossed product algebras to realize a larger algebra coinciding with the complete set of physical operators. In subsection \ref{sec: big crossed} we will formalize this construction. 

Let $H = H_1 \otimes H_2 \otimes H_3$ be a tripartite Hilbert space. Suppose that each $H_i$ is an internal frame for a common locally compact group $G$, so that each Hilbert space possesses an (over)complete basis of orientation states $\un{e}^{(i)}_g$ with compatible left $U^{(i)}$ and right $\overline{U^{(i)}}$ actions. Then, one can implement $G$ as a constraint group for the overall system $H$ by forming all different distributions of its kinematical degrees of freedom into internal frame and system. Each such choice will identify a different QRF. In particular, for any permutation $(ijk) \in S_3$ one can take $H_r = H_{k}$ for any $k$ leaving $H_s = H_{i} \otimes H_{j}$, or visa versa $H_r = H_{i} \otimes H_{j}$ leaving $H_s = H_k$. In other words, the set of QRFs are indexed by bi-partitions of the set $\{1,2,3\}$. Let us denote such a bi-partition as $(s \mid r)$ with $s$ referring to the selection of system degrees of freedom and $r$ a frame. 

The crossed product associated with the partition $(s \mid r)$ acts on the Hilbert space $H_{(s \mid r)} = H_s \otimes H_r$ with $H_s \equiv \bigotimes_{i \in s} H_i$ and $H_r \equiv \bigotimes_{i \in r} H_i$. A general state in $\Psi \in H_{(s \mid r)}$ can be written
\beq
	\Psi = \int_{G} \mu(g) \; f^{(s \mid r)}_{\Psi}(g) \otimes \un{e}^{(r)}_g, \qquad \un{e}^{(r)}_g \equiv \bigotimes_{i \in r} \un{e}^{(i)}_g, 
\eeq
where $f^{(s \mid r)}_{\Psi}$ is obtained by acting on $\Psi$ with Page-Wooters maps derived from the projectors of $\un{e}^{(r)}_g$.\footnote{We should note that in some examples considered in the literature \cite{ali_ahmad_quantum_2022} the orientation states resolve the identity only for a subspace $K_r \subset H_r$. In this case one should take $H_{(s \mid r)} = H_s \otimes K_r$. Nevertheless, this does not change the fact that $H_{(s \mid r)}$ is a covariant representation space and so the rest of the analysis goes through unchanged.} Acting on $H_{(s \mid r)}$ we can construct the representations $\lambda^{(s \mid r)}: G \rightarrow U(H_{(s \mid r)})$ and $\pi^{(s \mid r)}: B(H_s) \rightarrow B(H_{(s \mid r)})$ by direct analogy to \eqref{Frame induced unitary} and \eqref{Frame dressing}. To be precise:
\begin{flalign}
	&\lambda^{(s \mid r)}(g) \equiv \mathbb{1}_{H_s} \otimes U^{(r)}(g), \qquad U^{(r)}(g) \equiv \bigotimes_{i \in r} U^{(i)}(g), \nonumber \\
	&\pi^{(s \mid r)}(\mathcal{O}) \Psi \equiv \int_{G} \mu(g) \; \text{Ad}_{U^{(s)}(g)}(\mathcal{O}) f^{(s \mid r)}_{\Psi}(g) \otimes \un{e}^{(r)}_g, \qquad U^{(s)}(g) \equiv \bigotimes_{i \in s} U^{(i)}(g). 
\end{flalign}
The triple $(H_{(s \mid r)}, \pi^{(s \mid r)}, \lambda^{(s \mid r)})$ defines a covariant representation for the covariant system $(M_s \equiv B(H_s), G, \alpha^{(s \mid r)} \equiv \text{Ad}_{U^{(s)}})$. We emphasize here that the choice of QRF has also privileged a particular covariant system and in this respect the crossed products induced by different QRFs will not be trivially isomorphic. Nevertheless, the resulting crossed product from any \emph{fixed} QRF will be isomorphic to the canonical crossed product of its associated covariant system.  

Using the covariant representation on $H_{(s \mid r)}$ we define the relational crossed product:
\beq
	A_{(s \mid r)} \equiv \pi^{(s \mid r)}(M_s) \vee \lambda^{(s \mid r)}(G).  \label{Relational crossed product}
\eeq
All of the operators in $A_{(s \mid r)}$ are invariant under the automorphism $\theta^{(s \mid r)} \equiv \tilde{\alpha}^{(s \mid r)} \circ \text{Ad}_{\rho^{(r)}}$, with $\rho^{(r)}$ the right representation associated with the frame. Notice that each algebra $A_{(s \mid r)}$ is contained inside of the original kinematical algebra $B(H)$. This is because the dressed operators $\pi^{(s \mid r)}(M_s)$ are isomorphic to $B(H_s) \subset B(H)$ and the group representations $\lambda^{(s \mid r)}(G)$ come from the group representations $U^{(i)}(G)$ which are also inner relative to $B(H)$. Thus, one may interpret each $A_{(s \mid r)}$ as a \emph{distinct} subalgebra of physical operators under the constraint group $G$. Although the algebras associated with different system/frame divisions are generally distinct they may share some elements. For example, $A_{(12 \mid 3)}$ has a system algebra which is isomorphic to $B(H_1) \otimes B(H_2)$, but these operators are also contained in e.g., $A_{(1 \mid 23)}$, $A_{(2 \mid 31)}$, $A_{(23 \mid 1)}$, and $A_{(31 \mid 2)}$. Nevertheless, it is also clear that $A_{(12 \mid 3)}$ and $A_{(1 \mid 23)}$ do not share all of their elements and are therefore not isomorphic.

Roughly speaking, the full set of physical operators in $B(H)$ corresponds to the union of each $A_{(s \mid r)}$ modulo the intersections described above. Schematically, we \emph{define} this algebra by\footnote{Here $\mathcal{P}_2(3)$ is the set of all bi-partitions of $\{1,2,3\}$.}
\beq \label{Physical Algebra from Frames}
	A_{phys} \equiv \bigcup_{(s \mid r) \in \mathcal{P}_2(3)} A_{(s \mid r)}/\sim.
\eeq
The equivalence relation in \eqref{Physical Algebra from Frames} can be interpreted as encoding change of frame data relating physical operators in different reference frames whenever the resulting relational crossed products share isomorphic subalgebras. We shall describe these maps in detail now. We will then use these change of frame maps to address the problem of endowing the set $A_{phys}$ with an algebraic structure.  

By Landstad's construction, each $A_{(s \mid r)}$ fits into a sequence
\begin{equation} \label{Landstad sequence for relational crossed product}
\begin{tikzcd}
\mathcal{L}(G)
\arrow{r}{\lambda^{(s \mid r)}} 
& 
A_{(s \mid r)}
\arrow{r}{T^{(s \mid r)}} 
& 
M_{(s \mid r)},
\end{tikzcd}
\end{equation}
where $\lambda^{(s \mid r)}$ is a homomorphism and $T^{(s \mid r)}$ is an operator-valued weight. We can fill out this sequence by incorporating maps
\begin{equation} \label{Full Landstad}
\begin{tikzcd}
\mathcal{L}(G)
\arrow{r}{\lambda^{(s \mid r)}}  
& 
A_{(s \mid r)}
\arrow{r}{T^{(s \mid r)}} 
\arrow[bend left]{l}{\varpi^{(s \mid r)}}
& 
M_{(s \mid r)},
\arrow[bend left]{l}{\pi^{(s \mid r)}}
\arrow[bend left]{l} 
\end{tikzcd}
\end{equation}
with $\pi^{(s \mid r)}$ the dressing map and $\varpi^{(s \mid r)}$ is defined by\footnote{As we shall discuss, one can imagine a more general construction in which the maps of the lower sequence of \eqref{Full Landstad} are promoted to allow for the possibility that the sequence is not exactly split. This would be the analog of introducing a non-trivial group extension or, comparing to the extended phase space, having non-trivial curvature.}
\beq
	\varpi^{(s \mid r)} \circ \lambda^{(s \mid r)}(g) = g, \qquad \varpi^{(s \mid r)} \circ \pi^{(s \mid r)}(x) = e. 
\eeq

Now, suppose that two relational crossed products included in $A_{phys}$ admit subalgebras which are isomorphic to each other. In particular, let $U_{(s_1 \mid r_1)} \subset A_{(s_1 \mid r_1)}$ and $U_{(s_2 \mid r_2)} \subset A_{(s_2 \mid r_2)}$ be von Neumann algebras admitting isomorphisms $\Lambda_{i \rightarrow j}: U_{(s_i \mid r_i)} \rightarrow U_{(s_j \mid r_j)}$. Then we can put together the sequences restricted to these subalgebras to obtain the following commutative diagram:
\begin{equation}
\label{Change of Frame}
\begin{tikzcd}
&
\varpi^{(s_1 \mid r_1)}\bigg(U_{(s_1 \mid r_1)}\bigg)
\arrow[bend left]{r}{\lambda^{(s_1 \mid r_1)}}
\arrow{dd}{\Lambda^{(R)}_{1 \rightarrow 2}}
& 
U_{(s_1 \mid r_1)}
\arrow[left]{dd}{\Lambda_{1 \rightarrow 2}}
\arrow[left]{l}{\varpi^{(s_1 \mid r_1)}}
\arrow[bend left]{r}{T^{(s_1 \mid r_1)}}
& 
T^{(s_1 \mid r_1)}\bigg(U_{(s_1 \mid r_1)}\bigg)
\arrow[shift left]{l}{\pi^{(s_1 \mid r_1)}} 
\arrow{dd}{\Lambda^{(S)}_{1 \rightarrow 2}}
\\
&
&
&
&
\\
& 
\varpi^{(s_2 \mid r_2)}\bigg(U_{(s_2 \mid r_2)}\bigg)
\arrow[bend right, swap]{r}{\lambda^{(s_2 \mid r_2)}}
\arrow[shift left]{uu}{\Lambda^{(R)}_{2 \rightarrow 1}}
&
U_{(s_2 \mid r_2)}
\arrow[left, swap]{l}{\varpi^{(s_2 \mid r_2)}}
\arrow[bend right, swap]{r}{T^{(s_2 \mid r_2)}}
\arrow[shift left]{uu}{\Lambda_{2 \rightarrow 1}}
&
T^{(s_2 \mid r_2)}\bigg(U_{(s_2 \mid r_2)}\bigg)
\arrow[swap]{l}{\pi^{(s_2 \mid r_2)}} 
\arrow[shift left]{uu}{\Lambda^{(S)}_{2 \rightarrow 1}}
&
\end{tikzcd}
\end{equation}
Eqn. \eqref{Change of Frame} identifies change of frame maps relating system to system and frame to frame degrees of freedom:
\beq
	\Lambda^{(S)}_{i \rightarrow j} = T^{(s_j \mid r_j)} \circ \Lambda_{i \rightarrow j} \circ \pi^{(s_i \mid r_i)}, \qquad \Lambda^{(R)}_{i \rightarrow j} = \lambda^{(s_j \mid r_j)} \circ \Lambda_{i \rightarrow j} \circ \varpi^{(s_i \mid r_i)}. 
\eeq

As we have stressed, the change of frame map $\Lambda_{i \rightarrow j}$ is valid only on subalgebras of $A_{(s_i \mid r_i)}$ and $A_{(s_j \mid r_j)}$ which are isomorphic. To emphasize the physical interpretation of $A_{phys}$ and understand its underlying product structure it will be useful to define extended change of frame maps which are valid over the full algebras. Let $U^{ij}_{(s_i \mid r_i)}$ be the maximal subalgebra of $A_{(s_i \mid r_i)}$ `intersecting' with $A_{(s_j \mid r_j)}$ in the sense described above. Then, we define the following map $\Lambda^{ext}_{i \rightarrow j}: A_{(s_i \mid r_i)} \rightarrow A_{(s_j \mid r_J)}$,
\beq \label{Extended change of frame map}
    \Lambda^{ext}_{i \rightarrow j}(\mathfrak{X}) \equiv \begin{cases} 
      \Lambda_{i \rightarrow j}(\mathfrak{X}) & \mathfrak{X} \in U^{ij}_{(s_i \mid r_i)} \\
      \mathbb{1}_{A_{(s_j \mid r_j)}} & \mathfrak{X} \not\in U^{ij}_{(s_i \mid r_i)}. 
   \end{cases}
\eeq
The interpretation of \eqref{Extended change of frame map} is that it implements the isomorphism between $A_{(s_i \mid r_i)}$ and $A_{(s_j \mid r_j)}$ on `overlaps' and simply projects operators from $A_{(s_i \mid r_i)}$ which are non-overlapping with $A_{(s_j \mid r_j)}$ to the identity. Physically, this communicates the fact that if an operator lies completely outside of a given relational crossed product it is inaccessible to the observer whose QRF defines that algebra. 

We are now prepared to define a product structure on $A_{phys}$. Given $\mathfrak{X}, \mathfrak{Y} \in A_{phys}$ we first ask whether there exists a single relational crossed product algebra $A_{(s \mid r)}$ such that $\mathfrak{X},\mathfrak{Y} \in A_{(s \mid r)}$. Here we are regarding inclusion up to isomorphism induced by the change of frame maps \eqref{Change of Frame}. If such an algebra exists, we take the product $\mathfrak{X} \cdot \mathfrak{Y}$ simply to be the product between these elements regarded as operators in $A_{(s \mid r)}$. If such an algebra does not exist we take the product $\mathfrak{X} \cdot \mathfrak{Y} = \mathbb{1}$. In the former case we say that the operators $\mathfrak{X}$ and $\mathfrak{Y}$ are mutually observable. In the latter case, we say that they are not mutually observable. The product between non-mutually observable operators in $A_{phys}$ must be in the algebra $\mathbb{C} \mathbb{1}$; this coincides with the fact that such operators cannot be composed because they only exist in non-mutually consistent QRFs. 

Hopefully throughout this discussion it has become clear that the object $A_{phys}$ has the complexion of an algebraic manifold of some kind. The various crossed product algebras contained within this overarching algebra play the role of charts, with the change of frame maps \eqref{Change of Frame} informing the relationship between charts on mutually intersecting subsets. In this context we have been forced to carefully sort out more details than are present in the manifold case since we have to carry over algebraic data. Nevertheless, it has become clear that such an object is necessary to carefully encode all of the physical operators that are present in a system with numerous different QRFs. In the next subsection we formalize these observations into the definition of a novel algebraic object that encodes the data of all QRFs accessible in a given physical setting with constraint group $G$.

\subsection{The $G$-Framed algebra} \label{sec: big crossed}

Building upon the observations of the previous section, we are now prepared to make our definition of a $G$-framed algebra $\mathfrak{A}$. To do so we proceed in a series of steps that may be familiar from the construction of objects like manifolds and orbifolds \cite{caramello2022introduction}. 

Let $\mathfrak{A}$ be an involutive Banach algebra. A \emph{crossed product chart} for a subalgebra $A \subset \mathfrak{A}$ is a triple $(\tilde{A}, G, \phi)$ with $G$ a locally compact group, $\tilde{A}$ a $G$-crossed product algebra, and $\phi: \tilde{A} \rightarrow \mathfrak{A}$ a map that induces an isomorphism between $\tilde{A}$ and $A$. Given two crossed product charts $C_i \equiv (\tilde{A}_i, G_i, \phi_i)$, $i = 1,2$, an \emph{embedding} of $C_1$ into $C_2$ is an algebra inclusion $\lambda: \tilde{A}_1 \hookrightarrow \tilde{A}_2$ such that $\phi_1 = \phi_2 \circ \lambda$. Similarly, we say that the charts $C_1$ and $C_2$ \emph{overlap} if there exists a chart $C_{12} = (\tilde{A}_{12}, G_{12}, \phi_{12})$ which is embedded within both $C_1$ and $C_2$. 

A \emph{crossed product atlas} for $\mathfrak{A}$ is a collection of crossed product charts, $\mathcal{A} \equiv \{(\tilde{A}_i, G_i, \phi_i)\}_{i \in \mathcal{I}}$, that `cover' $\mathfrak{A}$ and are locally compatible. By cover we mean that $\mathfrak{A}$ is contained in the set union of $\phi_i(A_i)$:
\beq \label{coverage}
	\mathfrak{A} \subset \bigcup_{i \in \mathcal{I}} \phi_i(\tilde{A}_i). 
\eeq
The consideration of local compatibility recognizes that the union in \eqref{coverage} may overcount the elements in $\mathfrak{A}$ if the charts are overlapping. Thus, we require that for any two charts $C_i = (\tilde{A}_i, G_i, \phi_i)$, $i = 1,2$, and any element $\mathfrak{X} \in \mathfrak{A}$ which is contained (algebraically) in both $\phi_1(\tilde{A}_1)$ and $\phi_2(\tilde{A}_2)$, that there exists a third chart $C_{12}(\tilde{A}_{12},G_{12},\phi_{12})$ for which (a) $\mathfrak{X} \in \phi_{12}(\tilde{A}_{12})$ and (b) $C_{12}$ is mutually embedded within both $C_1$ and $C_2$. In other words, $C_{12}$ is an overlap between $C_1$ and $C_2$. 

Notice that the pair of embeddings $\lambda_{1}: C_{12} \hookrightarrow C_1$ and $\lambda_{2}: C_{12} \hookrightarrow C_2$ implicitly define an isomorphism $\Lambda_{1 \rightarrow 2}: \lambda_1(C_{12}) \rightarrow \lambda_2(C_{12})$. This is the change of frame map. We say that two charts are \emph{equivalent} if they can each be embedded into the other, and denote this equivalence by $\sim$. The algebra $\mathfrak{A}$ is equal to the union over the charts in a crossed product atlas modulo this equivalence:
\beq \label{G framed quotient}
	\mathfrak{A} = \bigcup_{i \in \mathcal{I}} \phi_i(\tilde{A}_i)/\sim. 
\eeq
An atlas $\mathcal{A}$ is called a \emph{refinement} of an atlas $\mathcal{B}$ if every chart in $\mathcal{A}$ admits an embedding of a chart in $\mathcal{B}$. Two atlases are deemed equivalent if they share a common refinement, and an atlas is called \emph{minimal} if it cannot be further refined. Finally, $\mathcal{A}$ is a $G$ atlas if, for each chart $C = (\tilde{A}, H, \phi) \in \mathcal{A}$ the group $H$ is a subgroup of $G$.

We can now introduce the $G$-framed algebra:
\begin{definition}[G-framed algebra]
	Given a locally compact group $G$ a $G$-framed algebra is an involutive Banach algebra $\mathfrak{A}$ along with an equivalence class of $G$-atlases. 
\end{definition}
Notice that a $G$-framed algebra whose minimal atlas consists of a single chart is nothing but a crossed product algebra. In this respect the $G$-framed algebra is a natural generalization of the crossed product. Relative to the discussion in Section \ref{sec: tripartite example} we see that $A_{phys}$ is a $G$-framed algebra in which each crossed product chart gives rise to a $G$-crossed product. More generally, the $G$-framed algebra can be regarded as the algebraic analog of an orbifold with individual crossed product charts coinciding with local trivializations therein. These local trivializations needn't be isomorphic, and may identify different symmetry subgroups $H \subset G$ or `system algebras' $M$ which should be interpreted as encoding the physics accessible to a local observer. In other words, each local chart encodes the QRF of a particular observer. 

It is profitable to think of a $G$-framed algebra as a global model for the quotient of a kinematical algebra by the action of the group $G$. In Appendix \ref{sec: Rieffel} we provide an alternative perspective on formulating such a quotient via Rieffel induction, which has been hypothesized to be a quantum analog of symplectic reduction. We provide some speculation on the relationship between Rieffel induction and the $G$-framed algebra, but a more rigorous exploration of this is left for follow up work.

As motivation for the construction of the $G$-framed algebra, let us recall the corresponding classical problem. What we will see is that the $G$-framed algebra addresses the same issues that are familiar from the point of view of the Gribov ambiguity. Let us be precise about what we mean here by considering the problem of symplectic reduction in a classical gauge theory. Recall that the extended phase space\footnote{A brief introduction to the extended phase space can be found in Appendix \ref{app: eps}. For a more complete introduction we refer the reader to \cite{Klinger:2023auu}.} $X_{ext}$ is a symplectic manifold with symplectic form $\Omega$ admitting a symplectomorphic action $R: G \times X_{ext} \rightarrow X_{ext}$ by the group $G$ which we deem to be a gauge redundancy. In a typical gauge theory this redundancy is encoded via a series of constraints $C_i: X_{ext} \rightarrow \mathbb{R}$, one for each generator of the Lie algebra of the group $G$, which are functions on $X_{ext}$ vanishing on physical phase space configurations. The constraint surface inside $X_{ext}$ is the locus of points where all of the constraints are mutually met. As we will discuss, this constraint surface will be identified with a quotient space $X_{ext}/G$ which generically must be described by an orbifold atlas -- the classical counterpart of the crossed product atlas for a $G$-framed algebra.  

Working in a local region $U \subset X_{ext}$ we can introduce a symplectic potential $\theta$ such that $\Omega = d\theta$. In \cite{Klinger:2023auu} it was demonstrated that there exists a canonical transformation which puts the symplectic potential into the form
\beq \label{Dressing symplectic potential}
\theta=\tilde\theta+\theta_G
\eeq
where 
\beq
\theta_G=\sum_i C_i\varpi^i
\eeq
with $\{C_i\}$ the constraints and $\varpi^i$ corresponding Maurer-Cartan forms, regarded as 1-forms on $X_{ext}$. This structure corresponds to the fact that the constraints generate the gauge symmetries on phase space. $\tilde\theta$ is a 1-form on the quotient space $U/G$ involving `dressed' variables which may be identified with orbits under the action $G$; choosing a particular representative for the form $\tilde\theta$ corresponds to a choice of gauge fixing. Implementing the constraints then reduces $\theta$ to $\tilde\theta$. 

A natural question to ask is how large we can make the region $U$ before the local description \eqref{Dressing symplectic potential} breaks down. Consider the tangent bundle $TX_{ext}$; then we may regard $\theta_G$ as defining a distribution ${\cal D}\subset TX_{ext}$. This means that $\theta_G$ pulls back to zero on ${\cal D}$.  Setting $n=\dim G$, the form $\wedge^n d\theta_G$, if non-zero, can be thought of as a volume form on the normal bundle and this will be true in all of $U$. If we try to extend the description \eqref{Dressing symplectic potential} beyond $U$, however, there may be points at which $\wedge^n d\theta_G=0$. At these points we conclude that two or more of the constraints have become linearly dependent. This will occur at points in the phase space where the group $G$ acts with non-trivial isotropy. Recall that the isotropy group at a point $x \in X_{ext}$ is given by
\beq
	G_x \equiv \{g \in G \; | \; R_g(x) = x\}.
\eeq
At such a point the description \eqref{Dressing symplectic potential} breaks down. Roughly speaking, there are less constraints being imposed upon the phase space than would naively be expected, and so the size of the quotient $X_{ext}/G$ at such a point will be \emph{larger} than is implied by the local quotient $U/G$. Instead, such a point should fit into a local chart $V \subset X_{ext}$ admitting a decomposition
\beq
	\theta^{(V)} = \tilde{\theta}^{(V)} + \theta^{(V)}_{H_V},
\eeq
where $\tilde{\theta}^{(V)}$ is a one form on the quotient space $V/H_V$ with $H_V \equiv G/G_{V}$ and $\theta^{(V)}_{H_V}$ consists of constraints associated with the active constraint group $H_V$. 

In general, the extended phase space can be covered by a series of charts defined by pairs $(U,H_U)$ where $U \subset X_{ext}$ is an open subset and $H_U = G/G_U$ is the quotient of the overall constraint group by the isotropy of the set $U$. In each chart the symplectic potential can be brought into the form
\beq
	\theta^{(U)} = \tilde{\theta}^{(U)} + \theta^{(U)}_{H_U},
\eeq
where $\tilde{\theta}^{(U)}$ identifies canonical pairs in the local quotient $U/H_U$, and $\theta^{(U)}_{H_U}$ collects the active constraints. Strictly speaking, this implies that the extended phase space has the structure of a $G$-bundle over a quotient space $X_{ext}/G$ which is rigorously described as an orbifold. The collection of charts $(U,H_U)$ define an orbifold atlas, provided they are constrained by appropriate conditions on overlaps \cite{caramello2022introduction}. In this sense, we see that the extended phase space possesses a global structure which closely mimics that of the $G$-framed algebra. Each local chart describes gauge-fixed physics in terms of a quotient space $U/H_U$ of dressed phase space fields. Heuristically, one may think of the charts of the $G$-framed algebra as emerging from a canonical quantization of these orbifold charts. 

The fact that $X_{ext}$ must be covered by multiple charts is a manifestation of what is usually referred to as the Gribov ambiguity. However, we should also note that the observation above goes beyond what is typically regarded as the Gribov problem. To understand this point, let us consider a special case in which $G$ acts freely on $X_{ext}$, i.e., the isotropy group $G_x = \{e\}$ for every $x \in X_{ext}$. In this case the quotient $X_{ext}/G$ is a manifold, and thus $X_{ext}$ can be regarded as a principal $G$-bundle. A different way of understanding this is that all of the constraints are active in each local chart and thus, although one requires multiple charts to sew together $X_{ext}$, a canonical transformation putting $\theta$ into the form \eqref{Dressing symplectic potential} is valid in every chart. In this sense, the canonical quantizations of each individual chart are all isomorphic, and at the level of the $G$-framed algebra these charts are treated as overlapping (i.e., extra copies are removed in the quotient \eqref{G framed quotient}). In other words, the minimal atlas for a $G$-framed algebra in this case would have a single chart, and the algebra would be interpreted as a crossed product. This is what one typically regards as a resolution to Gribov's problem; although one has multiple charts these are merely gauge copies and so it is sufficient to choose one when performing a canonical or path integral quantization. 

On the other hand, in the more general case where $G$ acts with non-trivial isotropy we see that this resolution fails. Different charts coincide with fundamentally different physical degrees of freedom and realize non-isomorphic quantizations. An example of this was given in \cite{vanrietvelde_switching_2021}, relative to the constraint quantization of an $N$-partite system in $3D$ space in which translations and rotations are treated as gauge symmetries. It is not hard to see that the constraint group acts with non-trivial isotropy in this case. A particularly stark example is when all $N$ particles are located at the origin. All rotations leave this phase space point invariant, and thus the only active constraints come from translations. At the end of the day there are six, five and three dimensional gauge orbits within the quotient of the overall phase space by the constraint group. In the $G$-framed algebra these distinct orbits would coincide with non-overlapping QRFs. In a general gauge theory, the lack of a single local chart which encodes, at least up to isomorphism, all of the physical configurations inside $X_{ext}/G$ necessitates having multiple charts to cover the full set of gauge invariant operators. Hence, in a typical gauge theory we expect that a single crossed product algebra will not be sufficient.

\subsection{A $G$-framed algebra for de Sitter space}\label{sec: de sitter}

To exemplify the structure of the $G$-framed algebra in a physically relevant situation, we turn to the case of semiclassical quantum gravity in the static patch of de Sitter. The relation between quantum reference frames and the de Sitter crossed product algebra has already been remarked in the literature~\cite{Fewster:2024pur,DeVuyst:2024pop}. We show here that the case of a static patch in de Sitter with multiple observers is described in the context of a G-framed algebra containing different crossed product algebras corresponding to the selection of each observer as a frame for the remaining degrees of freedom.

Suppose we are describing some matter quantum field theory and propagating gravitons in the static patch, where the algebras of observables of both fields acts on the Hilbert space $H_{\rm dS}$. We must impose the de Sitter isometry group as gauge constraints. The group gets broken down to the subgroup preserving the static patch, $G_{\rm P} \simeq \mathbb{R}_{t} \times G_{\rm compact}$ where the first group encodes time translations generated by $H_{\rm mod}$ and the other group is compact and encodes rotations. The invariant subalgebra under $G_{\rm P}$ is trivial~\cite{chandrasekaran_algebra_2022}, and so one fix would be to introduce a feature like an observer into the static patch to dress to in a gauge-invariant manner. In fact, we will introduce $N$ such observers each constituting a good QRF for $G_{\rm P}$. The ``kinematical" Hilbert space will then be~\cite{DeVuyst:2024pop}
\begin{equation}
    H_{kin} = H_{\rm dS} \otimes H_{i}^{\otimes_{i}^{N}}, \label{kinematical H space}
\end{equation}
where the second tensor factor is the tensor product of all the observer Hilbert spaces. The de Sitter constraints will relate the generators of the isometry transformations on each factor above. For simplicity, we will focus on the $\mathbb{R}_{t}$ subgroup of $G_{\rm P}$, which amounts to saying that we let our observers carry clocks instead of measurement devices for the full $G_{\rm P}$ group.

The Hamiltonian constraint will be 
\begin{equation}
    H_{\rm total} = H_{\rm mod} + \sum_{i}^{N} H_{i},
\end{equation} 
where $H_i$ is the $i$-th observer Hamiltonian. Choosing this observer to be the QRF amounts to selecting the system to be $ H_{s-i} = H_{\rm dS} \otimes H_{j}^{\otimes_{j\neq i}^{N}},$ and forming the non-canonical relational crossed product as in Eq.~\eqref{Relational crossed product}, which we refer to as $A_{i}$.

One can repeat this procedure for all $N$ observers to obtain a collection of relational crossed products labelled by the choice of observer, $\{A_{i}\}$. Given that each observer constitutes some representation of the isometry group, $A_{i} \not \simeq A_{j}$ generically. However, there are some shared operators between these local crossed products that must be identified. Fitting them all together under this relation yields the G-framed algebra of de Sitter, $\mathfrak{A}_{\rm dS}$. 

At this point, we should remark that so far this only treats the frames in $\mathfrak{A}$ as auxiliary systems adjoined in the process of taking the crossed product and then further projecting down to the chosen frame's unitary representation of $G$. In other words, $H_{s-i}$ for any $i$ has a fixed system $H_{\rm dS}$ shared by all dynamical systems appearing in the local crossed products of $\mathfrak{A}_{\rm dS}$. While each $A_{i}$ on its own is manifestly frame-dependent, there will exist a subalgebra $A_{\rm f.i}$ common to all $A_{i}$'s, which in this context will be frame-independent. However, we emphasize that this depends on what we mean by frame as one could have imagined including in $\mathfrak{A}_{\rm dS}$ frames that are built out of the matter quantum field theory for example, assuming they transform appropriately under $G$, thus potentially rendering $A_{\rm f.i}$ trivial. This, however, does not mean that there does not exist a notion of frame-independence as we discuss in Sec.~\ref{sec: rel states} and Sec.~\ref{sec: frame independence}.

One way to go beyond existing constructions in the literature~\cite{DeVuyst:2024pop} is to modify the previous discussion to account for multiple observers defining distinct, but partially overlapping static patches within de Sitter space. At the level of the kinematical Hilbert space in Eq.~\eqref{kinematical H space}, this amounts to allowing the de Sitter sector to be labelled by the choice of observer
\begin{equation}
    \tilde{H}_{kin} = \bigotimes_{i}^{N} H^{i}_{\rm dS} \otimes H_{i}.
\end{equation}
The fact that the multiple observers encoded in $\{H^{i}_{\rm dS} \otimes H_{i} \}$ define different but overlapping static patches and thus gravitational and field-theoretic degrees of freedom amounts to saying that there exists a subalgebra $M_{\rm overlap}$ common to all the algebras of observables of each static patch, $B(H^{i}_{\rm dS})$. While this takes further advantage of the richness of the $G$-framed algebra, we will return to an even more general discussion beyond a single semiclassical background below in Section~\ref{sec: semiclassical}.

\subsection{Relational density states and entropies} \label{sec: rel states}
What originally sparked interest in the crossed product in the physics community was Takesaki's theorem showing that the crossed product of a Type III$_1$ factor $M$ with its modular automorphism group is a semifinite von Neumann algebra $\hat{M}$~\cite{takesaki1973duality, witten_gravity_2022}. Shortly after, it was realized that the von Neumann entropy of semiclassical states of the modular crossed product is equivalent, up to an additive ambiguity, to the generalized entropy of the underlying subregion corresponding to the algebra~\cite{chandrasekaran_large_2022, Sorce:2023fdx, AliAhmad:2023etg}. 

Semifiniteness equips the algebra with a semifinite normal trace $\tau : \hat{M} \to \mathbb{C}$, which may be used to distinguish elements of $\hat{M}$ by whether or not they have finite trace. A modification of the dual weight theorem was presented in Ref.~\cite{Klinger:2023qna}, the upshot of which allows one to associate a state $\omega$ on $M$ with a density state $\rho^{\xi}_{\omega} \in \hat{M}$ which depends on how $M$ sits in $\hat{M}$ via an embedding $\xi$. 

Here, we remark that such an association is in principle always possible for a generic inclusion of von Neumann algebras as long as two conditions hold:
\begin{enumerate}
    \item There exists an embedding $\xi: M \to N$,
    \item The parent algebra $N$ is semifinite and so admits a trace $\tau$.
\end{enumerate}
In the context of quantum reference frames and the crossed product, we may derive the frame-dependence of objects like density states, entropies, and other calculable quantities using the above observation. Let $\mathfrak{A}$ be the G-framed algebra associated to some system of interest. Selecting a frame $i$ allows us to localize to a chart $A_{i} \subset \mathfrak{A}$, which acts on $H \otimes H_{r}$ where $H$ is a representation space of the associated ``system" algebra $M_{i}$ and $H_{r}$ is a Hilbert space carrying the relevant action of $G$ and encoding the $i$-th frame's degrees of freedom. Other choices of frames correspond to different crossed product charts $A_{j}$. While it is tempting to assume that $\mathfrak{A}$ itself is semifinite and apply the modified dual weight theorem to the inclusion $A_{i} \subset \mathfrak{A}$, it is not clear what notion of semifiniteness is applicable to $\mathfrak{A}$ given that it is generally only a Banach algebra. See Sec.~\ref{sec: frame independence} for more discussion on this.

The alternative route is to consider embeddings $\xi$ of  $M_{i} := M_{(\lambda, \delta)}$ into its associated crossed product  $A_{i}$, and assume that this latter algebra is semifinite with trace $\tau_{i}$ for some choice of frame $i$. In that case, given a state $\rho$ on $M_{i}$, one can induce a density state $\rho_{i}^{\xi} \in A_{i}$. This density state is manifestly frame-dependent; for a different choice of frame $j:= (\lambda', \delta')$, the system algebra itself generically changes to $M_{j} \not \simeq M_{i}$ for generic choices of $j$. If its associated crossed product is also semifinite with trace $\tau_{j}$, the induced density states of this algebra will house different different information about $\mathfrak{A}$ when compared to the original. To quantify this information, one could use the trace to compute quantum information quantities like the von Neumann entropy
\begin{equation}
    S_{vN, i}(\rho, \xi) = - \tau_{i} \left[\rho_{i}^{\xi} \ln \rho_{i}^{\xi} \right], 
\end{equation}
which is dependent on the choice of frame $i$ in addition to the usual dependence on the choice of embedding $\xi$.

It may seem that all physical objects one may consider are frame-dependent by definition, since a choice of frame $i$ selects out a different system $M_{i}$ and associated crossed product $A_{i}$. However, it is important to realize the non-triviality of the G-framed algebra $\mathfrak{A}$ is almost entirely contained in the quotient under change of reference frame maps. This allows information to seep across different subalgebras $A_{i}$ for different $i$, even if these correspond to inequivalent frames and descriptions. 

Let $A_{i}$ and $A_{j}$ correspond to two different local crossed product subalgebras arising from two different choices of frames $i$ and $j$ such that they share a non-trivial subalgebra $O_{ij}$. Moreover, assume, without loss of generality, that $A_{i}$ is semifinite so that there exists a trace $\tau_{i} : A_{i} \to \mathbb{C}$. Given that $O_{ij}$ is also contained in $A_{j}$, then we may at least assign a trace to some elements of $A_{j}$ which are contained in the intersection via $\tau_{i}(a_{ji})$ where $a_{ji} \in O_{ij}$. In other words, any $A_{j}$ sharing operators with a semifinite $A_{i}$ will be semifinite itself as long as elements in the overlap have finite trace. Then, observables sensitive to the intersection will encode information common to both choices of frames and any other frame sharing operators with either of them. This scenario arises in our previous example in de Sitter. The modular crossed product with a clock in the regular representation of $\mathbb{R}_{t}$ is semifinite and constitutes the subalgebra $A_{\rm reg} < \mathfrak{A}_{\rm dS}$. All $A_{i}$'s obtained from some restriction of the regular representation share operators with this subalgebra. These charts have the interpretation of choosing a different clock with a potentially different Hamiltonian as the frame. The entropy computed at the level of each chart is relational, and the frame-dependence is encoded in (1) the entropy of the chosen frame itself and (2) the relative entropies of the different systems one obtains by conditioning on the different choice of frame. In the context of the crossed product entropy being the same as generalized entropy of the underlying subregion, each $A_{i}$ would not necessarily agree on the `bulk matter' entropy part as the choice of frame changes what the system algebra is. Moreover, the frame itself contributes a term in the entropy which will differ as the frames may be described by inequivalent Hilbert spaces. This discussion is purely semiclassical, and we comment on using the G-framed algebra as a way to encode information beyond semiclassical gravity in Sec.~\ref{sec: semiclassical}.

Given that $\mathfrak{A}$ will have a complicated global structure due to the quotient, it will generically be the case that the intersection across all local crossed product algebras is trivial. In other words, not all framed algebras will share some operators. So, one cannot expect frame-independent information to arise from literal overlaps among \textit{all} relational crossed products. However, as in the case of a Riemannian manifold where all local charts must agree on curvature invariants, the quotient structure of $\mathfrak{A}$ ensures that some frame-independent information makes its way to the local crossed product subalgebras. One way to define frame-independent observables is to demand that they are invariant under all change of frame maps, but that might be too strict of a condition. We discuss some technical challenges in defining frame-independent notions and semifiniteness at the level of $\mathfrak{A}$ itself in Sec.~\ref{sec: frame independence}.
\section{Discussion} \label{sec: disc}
We outline some avenues for future work in the following.
\subsection{Frame-independent objects}\label{sec: frame independence}
In Sec.~\ref{sec: rel states}, we mainly discussed the relational aspects of the information contained in each local crossed product in the G-framed algebra. Since each such algebra is von Neumann, the familiar notions of semifiniteness in terms of existence of a semifinite tracial weight and the dual weight theorem allows us to make concrete statements about the frame-dependence or frame-independence of objects in these algebras. Here, we consider how these notions may manifest themselves at the level of the G-framed algebra by considering weights on $\mathfrak{A}$ itself.

We begin with the simple case where all the frames appearing in $\mathfrak{A}$ are equivalent, similar to the apparent Gribov ambiguity discussed above. In that case, Landstad's theorem implies that the G-framed algebra may be globally expressed as a von Neumann crossed product algebra. Geometrically, this corresponds to a globally trivial principal bundle which is a true manifold. Any observable in this algebra will necessarily be frame-independent, and if the algebra is semifinite, then all quantities computed using weights will be frame-independent as well. A natural generalization of this case would be when $\mathfrak{A}$ is not globally a crossed product but it is still some von Neumann algebra. In that case, the usual machinery applies. One can consider a weight $\omega : \mathfrak{A} \to \mathbb{C}$ and its restriction to each local crossed product $\omega_{i}: A_{i} \to \mathbb{C}$. Importantly, while $\omega_{i}$ may be a semifinite tracial weight on some of the subalgebras of $\mathfrak{A}$, it is not guaranteed that this holds for all the other subalgebras and by extension for $\mathfrak{A}$ itself. This just reflects the observation made above that there could be frame choices that are not sufficient for the regulation of the divergences of their system algebras within the same $\mathfrak{A}$.

Moving on to the slightly more complicated case, we assume that $\mathfrak{A}$ is only a C$^*$ algebra instead. From the structure theory of C$^*$ algebras, we know that $\mathfrak{A}$ may be concretely viewed as a subalgebra of bounded operators on some Hilbert space $H_{\mathfrak{A}}$. Since $\mathfrak{A}$ is built out of von Neumann algebras $A_{i}$ each represented on $H_{A_{i}} \simeq H_{M_{i}} \otimes H_{i}$ where the first factor is a representation Hilbert space for $M_{i}$, the system algebra relative to the choice of frame $H_{i}$, then we expect that $\bigoplus_{i} H_{A_{i}} \subseteq H_{\mathfrak{A}}$. In the case where there is no overlapping frames, we expect an equality of the two total Hilbert spaces. When the quotient in $\mathfrak{A}$ is non-trivial, then this relates the summands of the direct sum together in a non-trivial way reflecting the global topology of $\mathfrak{A}$. Any weight on $\mathfrak{A}$ may be represented as a vector state in its GNS representation, but each such weight will induce a weight on the $A_{i}$ subalgebras. In this case, one can view $\mathfrak{A}$ in a Hilbert space in which the framed description of the theory relative to the $i$-th frame arises by `tracing' out the other frames. The induced state on $A_{i}$ will generically be mixed even though the original state on $\mathfrak{A}$ is pure. This reflects the fact that $\mathfrak{A}$ houses everything there is to know about the system in a frame-independent manner, but choosing a frame generically amounts to a loss of information and the introduction of frame-dependent artifacts.

Finally, the most general case is where $\mathfrak{A}$ is only a Banach algebra. In this case, Hilbert representations of $\mathfrak{A}$ need not exist~\cite{berkson_representations_1969}. This reflects the fact that $\mathfrak{A}$, in the most general case, is an exotic algebraic object from a physical perspective. One must then search for its representations in a bigger class of spaces, namely Banach spaces~(cf.~\cite{khodsiani2015type}). If $\mathfrak{A}$ is represented on such a Banach space $E$, then nothing prohibits the Hilbert spaces of the local crossed products $A_{i}$ from being subspaces of $E$. If $E$ is strictly Banach, then this forbids us to define a global inner product that is suitable for all subalgebras $A_{i}$ \textit{and} respects the quotient under non-trivial overlaps. This suggests that this case has the interpretation of totally non-overlapping frames.

Given the above discussion, there are many technical questions one has to answer before saying anything concrete and rigorous about physics occurring at the level of $\mathfrak{A}$. However, it seems that the algebraic type of $\mathfrak{A}$ influences the interplay between the different frames and descriptions it houses. We hope to address  this more concretely in the future.
\subsection{Beyond semiclassical gravity?}\label{sec: semiclassical}
In the previous subsection, the issue of understanding how the Hilbert spaces of the local crossed products embed in the space $\mathfrak{A}$ is represented on was raised. In the case where there were some overlapping frames, we argued that the direct sum of the relational crossed product Hilbert spaces will have to be adjusted to respect the quotient structure of $\mathfrak{A}$. This suggests that there are non-trivial state overlaps and operator correlations across different charts of the G-framed algebra, which would have been trivial in the case of a direct sum Hilbert space. 

Here, we raise the possibility that the G-framed algebra, specifically its quotient structure, is a way to encode information about non-trivial quantum gravitational overlaps between distinct semiclassical backgrounds. One way to think about semiclassical quantum gravity from a canonical perspective is that the total Hilbert space is 
\begin{equation}
    \mathcal{H}^{s.c}_{\rm total} = \bigoplus_{i} \mathcal{H}_{i}, 
\end{equation}
where $\mathcal{H}_{i}$ is the Hilbert space of matter and propagating gravitons in the $i$-th semiclassical background $g_{i}$.\footnote{This point of view was recently explored in the context of de Sitter \cite{Witten:2023xze}.} This heuristic picture will only be true assuming one can safely ignore overlaps $\langle g_{i} | g_{j}\rangle$, which is accomplished for example by taking the $G_{\rm N} \to 0$ limit. This resembles the case of non-overlapping charts in the G-framed algebra. When the frames overlap, we lose the direct sum and it is subsumed by a non-factorizable Hilbert space.

This is to be contrasted with the de Sitter example that we have provided in Sec.~\ref{sec: de sitter}. In that case, as we stressed, there is a fixed kinematical algebra of observables $M_{\rm overlap}$ shared by all the different frames, namely the de Sitter algebra of matter and propagating gravitons in the overlap of static patches defined by the observers. This is the statement that we have \textit{fixed} the semiclassical background and are simply toggling between different observers when we traverse $\mathfrak{A}$.  We believe that the G-framed algebra is flexible enough to accommodate such a description of gravity where the quotient structure is instead informed by the overlaps between geometries that are obtainable from the gravitational path integral. In that case, traversing $\mathfrak{A}$ would not only switch between different observers but also different semiclassical backgrounds.

\subsection{Curvature and projectivity} \label{sec: curv}

The present work has been largely concerned with the consequences of non-trivial global topology relative to the operator algebras we use to describe gauge theories. The $G$-framed algebra is an immediate manifestation of this, with the appearance of incommensurate charts indicating an obstruction to the existence of a single global QRF in which all physical observables are present. With this being said, we have been largely agnostic to the origin and quantification of this obstruction. Building upon the sharp analogy between crossed product algebras and principal bundles it seems plausible that such obstructions could be described through an algebraic analog of curvature or holonomy. 

A natural place where `curvature' could present itself is in the sequence
\begin{equation} \label{Sequence2}
\begin{tikzcd}
\mathcal{L}(G)
\arrow{r}{\lambda} 
& 
A
\arrow{r}{T} 
& 
M.
\end{tikzcd}
\end{equation}
which appears in the top down specification of a crossed product algebra. In Section \ref{sec: tripartite example} we have introduced a splitting of \eqref{Sequence2} featuring the maps $\pi^{(s \mid r)}$ and $\varpi^{(s \mid r)}$ which encode the dressing of system degrees of freedom relative to a chosen QRF and the projection of operators in the crossed product into the group von Neumann algebra. In this note we have assumed, as is typical, that the map $\pi^{(s \mid r)}$ is a homomorphism which preserves the system algebra under dressing. Comparing the sequence \eqref{Sequence2} to the short exact sequence which defines an Atiyah Lie algebroid, 
\begin{equation} \label{Short Exact Sequence ALA 2}
\begin{tikzcd}
0
\arrow{r} 
& 
L
\arrow{r}{j} 
\arrow[bend left]{l} 
& 
A
\arrow{r}{\rho} 
\arrow[bend left]{l}{\omega}
& 
TX
\arrow{r} 
\arrow[bend left]{l}{\sigma}
&
0\,.
\arrow[bend left]{l} 
\end{tikzcd}
\end{equation}
the map $\pi^{(s \mid r)}$ can be interpreted as playing the role of a horizontal lift relative to the principal $G$-bundle associated with the algebroid $A$. In general, a horizontal lift is not expected to be a homomorphism of Lie brackets, and its failure in being one encodes the curvature of a horizontal distribution within the algebroid. 

Stated more plainly, the choice of horizontal lifting coincides with a choice of connection, and the connection encodes important data about the global topology of the algebroid. From this point of view, one may think of the dressing map $\pi^{(s \mid r)}$ as encoding an algebraic analog of a connection, in which case obstructions to this map being a homomorphism would encode important details about the topology of the algebra. In the $G$-framed algebra this data should be respected across charts which indicates that the algebraic connection and its associated curvature could encode interesting global, and potentially frame independent information -- not unlike characteristic classes. 

An alternative point of view on the splitting of the sequence \eqref{Sequence2} arises from the interpretation of the crossed product as a generalization of a group extension. Most physicists are familiar with this concept through the idea of a central extension. Given a group $G$ a central extension, $G_C$, is described by a short exact sequence
\begin{equation} \label{Central Extension}
\begin{tikzcd}
U(1)
\arrow{r}{} 
& 
G_C
\arrow{r}{} 
& 
G.
\end{tikzcd}
\end{equation}
The set of projective representations of $G$ are in one to one correspondence with splittings of the sequence \eqref{Central Extension}, which are maps $\sigma: G \rightarrow G_C$. The failure of the map $\sigma$ to be a homomorphism is encoded in the presence of a non-trivial two-cocycle $C: G \times G \rightarrow U(1)$ such that
\beq
    \sigma(g) \sigma(h) = C(g,h) \sigma(gh). 
\eeq
The map $C$ can be regarded as a cohomology class in a complex referred to as the Hochschild cohomology \cite{hochschild1947cohomology,kadison1971cohomology,raeburn2000twisted}. Each unique group cohomology class encodes a distinct projective represenation. From this perspective, non-trivial splittings of the sequence \eqref{Sequence2} can be interpreted as encoding generalized representations of the system algebra as it is embedding in the crossed product. In other words, non-trivial dressings of the system to the associated QRF. Previous work \cite{sutherland1980cohomology,sutherland1980cohomologyII,landstad1987representations} has explored a similar point of view relative to the sequence
\begin{equation} \label{Sutherland Sequence}
\begin{tikzcd}
M
\arrow{r}{} 
& 
A
\arrow{r}{} 
& 
\mathcal{L}(G),
\end{tikzcd}
\end{equation}
in which case they have termed the central algebra a \emph{twisted crossed product}. This algebra is realized by following the same steps as one would to obtain a standard crossed product, only relaxing the map $\lambda^{(K)}$ in the covariant representation $(K,\pi^{(K)}_{\alpha},\lambda^{(K)})$ from a representation to a projective representation. In future work we intend to investigate how these and other related twistings can be used to classify $G$-framed algebras according to their topological invariants, and the relationship between these invariants and the frame independent quantities described above.

\appendix
\renewcommand{\theequation}{\thesection.\arabic{equation}}
\setcounter{equation}{0}

\section*{Acknowledgments}
S.A.A would like to thank Alexander R.~H.~ Smith for discussions on quantum reference frames and the algebraic approach to quantum theory and Ahmed Almheiri, Ro Jefferson, and Simon Lin for discussions on the crossed product. M.S.K. would like to thank Michael Stone for various discussions about von Neumann algebras and especially crossed products. The authors thank Philipp H\"{o}hn for his insightful comments on the draft. R.G.L. thanks the Perimeter Institute for Theoretical Physics for support, and acknowledges the U.S. Dept. of Energy grant DE-SC0015655.

\section{The Extended Phase Space} \label{app: eps}

In this appendix we consider the symplectic geometry associated with a gauge theory in terms of the extended phase space. Following the analysis of \cite{Klinger:2023tgi}, one may regard the following as a classical analogue of a crossed product. Along the way, we will see how the extended phase space provides a natural geometric setting for the relational formalism (in its classical form), which has also been discussed in \cite{Carrozza_2022,carrozza2023edgemodesdynamicalframes,goeller2022diffeomorphisminvariantobservablesdynamicalframes,Kabel_2023}. 

Let $(X,\omega)$ be a symplectic manifold, and denote by $M^{pq}_{X}$ its associated Poisson algebra. That is $M^{pq}_{X}$ consists of functions $f: X \rightarrow \mathbb{R}$ along with a bracket induced by the symplectic form $\omega$ in the following way: To each function $f \in M^{pq}_{X}$ we can associate a vector field $\un{V}_f \in TX$ solving
\beq \label{Hamiltonian vector field}
	df + i_{\un{V}_f} \omega = 0. 
\eeq
The integral curves generated by $\un{V}_f$ are referred to as the Hamiltonian flow induced by $f$, and $\un{V}_f$ is called a Hamiltonian vector field. The Poisson bracket is given by
\beq \label{Poisson bracket}
	\{f,h\}_{M^{pq}_{X}} \equiv -i_{\un{V}_f} i_{\un{V}_h} \omega.
\eeq
Using \eqref{Hamiltonian vector field} we can rewrite \eqref{Poisson bracket} as
\beq
	\{f,h\}_{M^{pq}_{X}} = i_{\un{V}_f} dh = \mathcal{L}_{\un{V}_f} h,
\eeq
In other words, the Poisson bracket of $f$ and $h$ is nothing but the Lie derivative of $h$ along the Hamiltonian flow induced by $f$. 

For our purposes we will be interested in the case that the symplectic manifold $(X,\omega)$ admits a $G$ action $a: G \times X \rightarrow X$. We say that this action is \emph{symplectomorphic} if it preserves the symplectic form as
\beq \label{Symplectomorphism}
	a_g^* \omega = \omega, \; \forall g \in G. 
\eeq
A symplectomorphic action induces a homomorphism of the Poisson algebra via pullback. That is, $f \mapsto a_g^*f$. Using \eqref{Symplectomorphism} we can write\footnote{Here we have used the fact that for an invertible map $\phi: X \rightarrow X$,
\beq
	\phi^*\bigg(i_{\phi_*\un{V}} \omega\bigg) = i_{\un{V}}\phi^*\omega,
\eeq
which further implies
\beq
	i_{\phi_* \un{V}} \omega = (\phi^{-1})^*\bigg(i_{\un{V}}\phi^*\omega\bigg).
\eeq}
\beq
	i_{(a_{g^{-1}})_*\un{V}_f} \omega = a_g^*\bigg(i_{\un{V}_f}\omega\bigg). 
\eeq	
From which we conclude that
\beq
	0 = a_g^*\bigg(df + i_{\un{V}_f} \omega\bigg) = d a_g^*f + i_{(a_{g^{-1}})_* \un{V}_f} \omega. 
\eeq
In other words, under the mapping $f \mapsto a_g^*f$ the associated Hamiltonian vector field is mapped as $\un{V}_f \mapsto (a_{g^{-1}})_*\un{V}_f$. Thus, we have
\beq
	\{a_g^*f, a_g^*h\}_{M^{pq}_{X}} = -i_{\un{V}_{a_g^*f}}i_{\un{V}_{a_g^*h}} \omega = a_g^*\bigg(-i_{\un{V}_f} i_{\un{V}_h} \omega \bigg) = a_g^*\bigg(\{f,h\}_{M^{pq}_{X}}\bigg).
\eeq 
The triple $(M^{pq}_{X}, G, a^*)$ should be thought of as the symplectic analog of a covariant system $(M,G,\alpha)$ as introduced in Section \ref{sec: bottom up}.

We will hereafter work under the assumption that the action $a$ is symplectomorphic. Thus, for each $g \in G$ the map $a_g: X \rightarrow X$ is a symplectomorphism. These maps are infinitesimally generated by the integral curves of vector fields $\xi_{\un{\mu}} \in TX$ where $\un{\mu} \in \mathfrak{g}$ is a Lie algebra element integrating to the desired group element. More rigorously,
\beq
	\xi_{\un{\mu}} \equiv \bigg(a_{\text{exp}(t\un{\mu})}\bigg)_* \frac{d}{dt},
\eeq
where $\text{exp}: \mathfrak{g} \rightarrow G$ is the standard exponential map. As $a_g$ is a symplectomorphism for each $g \in G$ it is immediately clear that $\xi_{\un{\mu}}$ will be a symplectic vector field for each $\un{\mu} \in \mathfrak{g}$:
\beq
	\mathcal{L}_{\xi_{\un{\mu}}} \omega = 0.
\eeq
However, it is not in general true that each $\xi_{\un{\mu}}$ is Hamiltonian in the sense that they generate the Hamiltonian flow of an element of the Poisson algebra. In other words, although the group $G$ acts \emph{on} $M^{pq}_{X}$ it is not immediately clear that the group $G$ can be regarded as \emph{in} $M^{pq}_{X}$. 

In previous work we have introduced an approach to augmenting the symplectic geometry $(X,\omega)$ such that the action $a$ can always be promoted to a Hamiltonian (and moreover equivariant) action on the resulting `extended phase space' \cite{Klinger:2023qna}. That is, given a covariant system $(M^{pq}_{X},G,a^*)$ there exists an extended Poisson algebra $M^{pq}_{X_{ext}}$ for which the group $G$ may be regarded as \emph{inner}. Of course, this is evocative of the crossed product construction discussed in the main text. The extended phase space is formally realized as a principal bundle $X_{ext} \rightarrow X$ with structure group $G$.\footnote{Strictly speaking this is a simplified case in which the action of $G$ is assumed to be free. More generally, as has been discussed in Section \ref{sec: big crossed}, the extended phase space $X_{ext}$ should be regarded as a $G$-bundle over a quotient $X_{ext}/G$ which has the structure of an orbifold. In this case the local charts of the extended phase space are non-isomorphic, corresponding to distinct crossed product charts in the $G$-framed algebra.} The structure maps of the principal bundle are given as
\beq \label{Structure maps for Xext}
	\pi: X_{ext} \rightarrow X, \; (g,x) \mapsto a_g(x), \qquad R: G \times X_{ext} \rightarrow X_{ext}, \; R_h(g,x) = (gh, a_{h^{-1}}(x)). 
\eeq
These maps are compatible in the sense that $\pi \circ R_h = \pi, \; \forall h \in G$. 

In eqn. \eqref{Structure maps for Xext}, we have tacitly presented $X_{ext}$ in a locally trivialized form. For the purposes of the present note it will be sufficient to treat the extended phase space in a local trivialization whereupon it is of the form of a product space $X_{ext} \simeq X \times G$. In other words, the effect of extending the phase space may simply be interpreted as attaching to the unextended system a series of probes valued in the group $G$. As we shall see, the inclusion of these extended degrees of freedom realize a relational frame, or, in other words, an `observer'. With this being said, understanding the physical interpretation of non-trivial topology in $X_{ext}$ will be an important part of future work in this subject. We will have more to say about this point in Section \ref{sec: disc}. 

For a more technical construction of the extended phase space we refer the reader to \cite{Klinger:2023tgi,Klinger:2023qna}.\footnote{In those notes we stress the role of the Atiyah Lie algebroid \cite{Ciambelli:2021ujl,Jia:2023tki} in naturally formulating the extended phase space, as opposed to symplectic manifold oriented approach used here.} In this note, it will be sufficient to restrict our attention to the resulting Poisson algebra $M^{pq}_{X_{ext}}$. Briefly, moving to the extended phase space involves two important promotions -- first the symplectic potential\footnote{E.g. the one form $\theta \in \Omega^1(X)$ for which $\omega = d\theta$. Strictly speaking $\omega$ is only locally exact, but we are only working in a local trivialization so this is sufficient.} $\theta \in \Omega^1(X)$ is promoted to the extended symplectic potential $\theta^{ext} \in \Omega^1(X_{ext})$ and secondly the infinitesimal generators of the group action are promoted from $\xi_{\un{\mu}} \in TX$ to $\xi^{ext}_{\un{\mu}} \in TX_{ext}$. In fact the promotion of $\xi_{\un{\mu}} \mapsto \xi^{ext}_{\un{\mu}}$ is induced by the promotion of the action $a: G \times X \rightarrow X$ to the right action $R: G \times X_{ext} \rightarrow X_{ext}$. In this way, we can read off the extended symmetry generating vector fields as
\beq \label{Extended Hamiltonian Vector Fields}
	\xi^{ext}_{\un{\mu}} = -\xi_{\un{\mu}} \oplus \un{\mu},
\eeq
where here $\un{\mu}$ is regarded as a vector field on $G$ via the identification $\mathfrak{g} \simeq T_e G$. 

Thus, it remains only to specify the extended symplectic potential which is fixed by demanding that\footnote{Hereafter we use hats to distinguish geometric operators on $X_{ext}$ from those on $X$.}
\beq \label{Equivariance}
	\hat{\mathcal{L}}_{\xi^{ext}_{\un{\mu}}} \theta^{ext} = 0,
\eeq
and that the symplectic potential agrees with $\theta$ when contracted with vector fields on the base $X$. An immediate corollary of \eqref{Equivariance} is that the map
\beq
	\Phi: \mathfrak{g} \rightarrow M^{pq}_{X_{ext}}, \; \un{\mu} \mapsto \hat{i}_{\xi^{ext}_{\un{\mu}}} \theta^{ext}
\eeq
defines a Hamiltonian function whose Hamiltonian flow is generated by $\xi^{ext}_{\un{\mu}}$ for each $\un{\mu} \in \mathfrak{g}$. In fact, it moreover implies that the set of functions $\Phi_{\un{\mu}}$ form a representation of the Lie algebra embedded inside the Poisson algebra
\beq \label{GG Bracket}
	\{\Phi_{\un{\mu}}, \Phi_{\un{\nu}}\}_{M^{pq}_{X_{ext}}} = \Phi_{[\un{\mu},\un{\nu}]_{\mathfrak{g}}}. 
\eeq
The second consideration implies that the Poisson algebra $M^{pq}_{X}$ can be regarded as a subalgebra of $M^{pq}_{X_{ext}}$ by naive inclusion -- that is functions $f,h: X \rightarrow \mathbb{R}$ are treated as functions on $X_{ext}$ and their Poisson algebra is preserved:
\beq \label{XX Bracket}
	\{f,h\}_{M^{pq}_{X_{ext}}} = \{f,h\}_{M^{pq}_{X}}. 
\eeq
When combined, these two conditions imply that the functions $\Phi_{\un{\mu}}$ implement a (infinitesimal) representation of the action $a^*$ via the Poisson bracket:
\beq \label{GX Bracket}
	\{\Phi_{\un{\mu}},f\}_{M^{pq}_{X_{ext}}} = -\mathcal{L}_{\xi_{\un{\mu}}} f. 
\eeq

Together eqns. \eqref{GG Bracket}, \eqref{XX Bracket}, and \eqref{GX Bracket} indicate that the full Poisson algebra $M^{pq}_{X_{ext}}$ is roughly of the form $M^{pq}_{X} \oplus C^*(G)$, where here $C^*(G)$ is the group algebra associated with $G$. We can now confront the question of implementing constraints from the perspective of the extended phase space. An element $\mathfrak{F} \in M^{pq}_{X_{ext}}$ is $G$ invariant if it is invariant under the action of the pullback $R^*: G \times M^{pq}_{X_{ext}} \rightarrow M^{pq}_{X_{ext}}$. Of course, if $R_g^*\mathfrak{F} = \mathfrak{F}$ for each $g \in G$ it will also be true that
\beq \label{Invariant -> Commutation}
	\{\Phi_{\un{\mu}}, \mathfrak{F}\}_{M^{pq}_{X_{ext}}} = \hat{\mathcal{L}}_{\xi^{ext}_{\un{\mu}}} \mathfrak{F} = 0, \; \forall \un{\mu} \in \mathfrak{g}.  
\eeq
Thus, in terms of the extended action $R$ we can realize a condition for invariance in terms of the Poisson commutation of the element $\mathfrak{F}$ with all of the Hamiltonian functions generating the constraints. 

A natural set of invariant observables is obtained by `dressing' ordinary observables $f \in M^{pq}_{X}$. Let
\beq	 \label{Dressing}
	\pi_{a}: M^{pq}_{X} \rightarrow M^{pq}_{X_{ext}}, \; \bigg(\pi_a(f)\bigg)(g,x) \equiv f \circ a_g(x). 
\eeq
Then, it is straightforward to see that $R_h^* \pi_a(f) = \pi_a(f)$ since $G$ now acts on $\pi_a(f)$ in two compensating ways:
\beq \label{Invariance of Dressed Observables}
	\bigg(R_h^* \pi_a(f)\bigg)(g,x) = \bigg(\pi_a(f)\bigg)(gh, a_{h^{-1}}(x)) = f \circ a_{gh}(a_{h^{-1}}(x)) = \bigg(\pi_a(f)\bigg)(g,x). 
\eeq
Thus, the set $\pi_a(M^{pq}_{X}) \subset M^{pq}_{X_{ext}}$ is in fact an invariant subalgebra under the action $R$ with
\beq \label{Invariance of Dressed Observables 2}
	\{\Phi_{\un{\mu}}, \pi_a(f)\}_{M^{pq}_{X_{ext}}} = 0, \; \forall f \in M^{pq}_{X}, \un{\mu} \in \mathfrak{g}. 
\eeq
The observations made in \eqref{Dressing}-\eqref{Invariance of Dressed Observables 2} may be interested as the classical analog of the commutation theorem discussin in Section \ref{sec: bottom up}.


In comparison to dressing \eqref{Dressing}, there is an alternative but closely related approach to implementing the constraints which comes in the form of gauge fixing. Instead of mapping $f$ into an orbit of observables, we can choose a single representative of its orbit and map every member to that representative. In \cite{Klinger:2023auu} it has been shown how gauge fixing can be understood in terms of a map $T_{\mathcal{F}}: M^{pq}_{X_{ext}} \rightarrow M^{pq}_{X}$ formulated as a Faddeev-Popov integral. The map $T_{\mathcal{F}}$ is defined in terms of a gauge non-invariant function, $\mathcal{F}: X \rightarrow \mathfrak{g}$, whose kernel intersects each $G$-orbit exactly once. That is,
\beq \label{Gauge Fixing Functional}
	\forall x \in X \; \exists! g \in G \; \text{s.t. } \mathcal{F} \circ a_g(x) = 0. 
\eeq
We denote the unique solution to \eqref{Gauge Fixing Functional} for a given $x \in X$ by $z_{\mathcal{F}}(x) \in G$, that is $\mathcal{F} \circ a_{z_{\mathcal{F}}(x)}(x) = 0$. Let $a_G(x) \equiv \{a_g(x) \in X \; | \; g \in G\}$ denote the gauge orbit of $x \in X$. Notice that,
\beq \label{Invariance of representative}
	a_{z_{\mathcal{F}}(x_1)}(x_1) = a_{z_{\mathcal{F}}(x_2)}(x_2), \; \forall x_1, x_2 \in a_G(x).
\eeq
That is the assignment $x \mapsto a_{z_{\mathcal{F}}(x)}(x) \equiv [x]_{\mathcal{F}}$ defines a unique representative of each gauge orbit. Then, we can define the integral\footnote{Here, $\delta(\mathcal{F} \circ a_g(x))$ is a normalized delta function. To correctly obtain the normalization, one must construct a Faddeev-Popov determinant.}
\beq
	\bigg(T_{\mathcal{F}}(\mathfrak{F})\bigg)(x) \equiv \int_{G} \mu(g) \; \delta\bigg(\mathcal{F} \circ a_g(x)\bigg) \mathfrak{F}(g,x) = \mathfrak{F}(z_{\mathcal{F}}(x),x). 
\eeq 
In particular, we see that
\beq \label{Gauge Fixed Observables}
	\bigg(T_{\mathcal{F}}(\pi_a(f))\bigg)(x) = f \circ a_{z_{\mathcal{F}}(x)}(x) = f([x]_{\mathcal{F}})
\eeq
depends only on the representative of the gauge orbit. Observables as obtained from \eqref{Gauge Fixed Observables} are immediately $G$-invariant --
\beq
	R_g^*\bigg(T_{\mathcal{F}}(\pi_a(f))\bigg)(x) = a_g^*\bigg(T_{\mathcal{F}}(\pi_a(f))\bigg)(x) = f \circ a_{z_{\mathcal{F}} \circ a_g(x)}(a_g(x)) = f([x]_{\mathcal{F}}),
\eeq
where we have used \eqref{Invariance of representative} with $x_1 = x$ and $x_2 = a_g(x)$.

The gauge fixing approach \eqref{Gauge Fixed Observables} may be understood in terms of conditionalization in the relational formalism. Indeed, 
\beq \label{Gauge Invariant Extension}
	T_{\mathcal{F} - \un{\mu}}(\pi_a(f)) = F_{f,\mathcal{F}}(\un{\mu}),
\eeq 
where here $F_{f,\mathcal{F}}(\un{\mu})$ is the `gauge invariant extension of gauge-fixed quantity' as defined in \cite{hoehn_trinity_2021}. Here, we have taken a slightly modified gauge fixing functional in which $\mathcal{F}$ is not set equal to zero, but rather is fixed to a constant value $\un{\mu} \in \mathfrak{g}$.\footnote{This is the generalization of the relational formalism when there is more than one constraint. In the Trinity paper, for example, $\un{\mu} = \tau$ is a `time' variable.} In words \eqref{Gauge Invariant Extension} is the observable defined by $f \in M^{pq}_{X}$ conditional on $\mathcal{F} - \un{\mu} = 0$, or simply the observable obtained from $f$ when $\mathcal{F} = \un{\mu}$. In the case where there is a single constraint, $\mathcal{F}$ can be regarded as a clock function, and $\un{\mu} \sim \tau$ as the internal time read by this clock. Then, \eqref{Gauge Invariant Extension} defines the gauge invariant observable obtained from $f$ when the clock reads internal time $\tau$. 

An important observation in \eqref{Gauge Invariant Extension} is that it leaves behind a residual freedom in terms of the choice of gauge fixing functional; altering this choice will change the representative of each gauge orbit. Working with gauge fixing functionals of the form $\mathcal{F} - \un{\mu}$ we can absorb this freedom into the right to change $\un{\mu}$. Thus, $F_{f,\mathcal{F}}(\un{\mu})$ should be regarded as a $\mathfrak{g}$-parameterized family of gauge invariant observables. Allowing $F_{f,\mathcal{F}}(\un{\mu})$ to `flow' defines the relational evolution of $f$ with respect to the `observer' defined by $\mathcal{F}$. 

\begin{table}[h!]
\centering
\begin{tabular}{ll}
\textbf{Extended Phase Space}                  & \textbf{Relational Formalism}                   \\
$X_{ext} \sim X \times G$, Kinematical Phase Space              & $P_{kin} \sim P_{sys} \times P_{clock}$, Kinematical Phase Space \\
$\Phi_{\un{\mu}} \in M^{pq}_{X_{ext}}$, Constraint Hamiltonians & $C = H_{sys} + H_{clock}$, Constraint Hamiltonian                \\
$\bigg(\pi_a(f)\bigg)(g,x) \equiv f \circ a_g(x)$, Dressed observable               & Clock (Observer) Neutral Observables                             \\
$T_{\mathcal{F} - \un{\mu}}(\pi_a(f))$, Gauge-fixed observable  & $F_{f,\mathcal{F}}(\un{\mu})$, Relational observable             \\
$\mathcal{F}$, Gauge-fixing functional                          & Clock/Observer                                                  
\end{tabular}
\caption{Dictionary relation classical symplectic analysis of extended phase space and the classical relational formalism. }
\label{Extended Phase Space vs Classical Relational Formalism}
\end{table}

In table \ref{Extended Phase Space vs Classical Relational Formalism} we have provided an overview of the correspondence between the extended phase space and the relational formalism in the classical context.\footnote{Notice that the constraint Hamiltonians, $\Phi_{\un{\mu}}$, do not necessary split as the sum of two terms in the extended phase space. Nevertheless, the vector field generating the extended $G$ action, $R: G \times X_{ext} \rightarrow X_{ext}$ has the form \eqref{Extended Hamiltonian Vector Fields} which is more reminiscent of $C = H_{sys} + H_{clock}$.}

\section{Group von Neumann algebra} \label{app: group vN}

In this appendix we review the construction of the group von Neumann algebra. Let $G$ be a locally compact group and denote by $\ell: G \rightarrow L^2(G)$ the left regular representation of $G$ on $L^2(G)$. The group von Neumann algebra is the von Neumann algebra obtained by closing the aforementioned representation in the weak operator topology induced by $L^2(G)$: $\mathcal{L}(G) \equiv \ell(G)''$. Alternatively, the group von Neumann algebra can be obtained as follows: Let $C_0(G)$ denote the space of continuous and compactly supported functions on $G$. $C_0(G)$ can be turned into an involutive Banach algebra by introducing the following product and involution:
\beq \label{C0 algebra}
	\eta \star \zeta(g) \equiv \int_{G} \mu(h) \eta(h^{-1}g) \zeta(h), \qquad \eta^*(g) = \delta(g^{-1}) \overline{\eta(g^{-1})}, \; \eta,\zeta \in C_0(G).
\eeq
The algebra $C_0(G)$ possesses a $*$-representation on $L^2(G)$, $c: C_0(G) \rightarrow B(L^2(G))$,
\beq
	c(\eta)(f)(g) = \int_{G} \mu(h) \eta(h^{-1}g) f(h). 
\eeq
The group von Neumann algebra may then equivalently be realized as the closure $\mathcal{L}(G) = c(C_0(G))''$. From the latter point of view, we automatically obtain a faithful, semi-finite, normal weight on $\mathcal{L}(G)$ in terms of the inner product on $L^2(G)$:
\beq \label{Plancherel Weight}
	\gamma: \mathcal{L}(G) \rightarrow \mathbb{C}, \qquad \gamma(\eta^* \star \zeta) = g_{L^2(G)}(\eta, \zeta) = \int_{G} \mu(h) \overline{\eta(h)} \zeta(h) = \eta^* \star \zeta(e). 
\eeq
The weight defined in \eqref{Plancherel Weight} is called the Plancherel weight of the group $G$.

\section{Comparison between $G$-framed algebra and Rieffel Induction} \label{sec: Rieffel}

In Section \ref{sec: big crossed} we introduced the $G$-framed algebra and argued that it should be regarded as an algebraic analog for the global quotient of a manifold with a locally compact group. The appearance of multiple quantum reference frames within the $G$-framed algebra is subsequently interpreted as a manifestation of the non-trivial topology of the resulting quotient, which generically may only be covered by a series of local charts.  In this section we review Rieffel induction \cite{rieffel1974induced} and discuss Landsman's interpretation \cite{landsman1995rieffel} of Rieffel induction as a quantum version of Marsden-Weinstein reduction \cite{marsden1974reduction}. Marsden-Weinstein is an approach to implementing quotients of symplectic manifolds by symmetry groups. Thus, one should expect the quantum analog of this procedure to realize an algebraic quotient of a similar complexion to the $G$-framed algebra. As we shall see, this is the case and the resulting algebraic object possesses a similar `framed' description with isomorphism implemented by concept of imprimitivity. 

\subsection{Rieffel Induction}

Let $A$ and $B$ be a pair of $C^*$ algebras. A $B$-rigged space is a Banach space $X$ admitting a right representation $r_B: B \rightarrow B(X)$ along with a $B$-valued inner product $G_B: X \times X \rightarrow B$ compatible with the $B$-representation in the sense that
\beq \label{B valued inner}
	G_B(x_1, r_B(b) x_2) = G_B(x_1,x_2) b, \; \forall x_1,x_2 \in X, \; b \in B. 
\eeq
Morally, a $B$-rigged space is a cousin of a Hilbert space in which elements in the $C^*$ algebra $B$ are treated as `scalars'. If $G_B(x,x) = 0 \iff x = 0$ we say that $G_B$ is definite. 

The set of bounded operators on $X$ viewed as a $B$-rigged space consists of linear maps $\mathcal{O}: X \rightarrow X$ satisfying the following conditions: 
\begin{enumerate}
	\item There exists a constant $k > 0$ such that\footnote{Here inequality is in the sense appropriate to the algebra $B$.}
\beq
	G_B(\mathcal{O}(x), \mathcal{O}(x)) \leq k^2 G_B(x,x), \; forall x \in X,
\eeq
	\item There exists a map $\mathcal{O}^{\dagger}: X \rightarrow X$ satisfying $(1)$ above and for which
	\beq
		G_B(\mathcal{O}(x_1),x_2) = G_B(x_1, \mathcal{O}^{\dagger}(x_2)), \; \forall x_1, x_2 \in X, 
	\eeq
	\item As a map $\mathcal{O}$ commutes with the right action $r_B$. 
\end{enumerate}
We shall denote the set of bounded operators on $X$ relative to the $B$-rigging $G_B$ as $B_B(X)$. In the event that $G_B$ is definite, the adjoint $\mathcal{O}^{\dagger}$ is uniquely defined. Moreover, every operator satisfying $(1)$ and $(2)$ above will automatically satisfy $(3)$. This follows from a simple computation:
\beq
	G_B(x_1, \mathcal{O} \circ r_B(b) x_2) = G_B(x_1, \mathcal{O}x_2)b = G_B(x_1, r_B(b) \circ \mathcal{O} x_2),
\eeq
or in other words,
\beq
	G_B(x_1, \bigg(\mathcal{O} \circ r_B(b) - r_B(b) \circ \mathcal{O}\bigg) x_2) = 0, \; \forall x_1, x_2 \in X, b \in B. 
\eeq
If $G_B$ is definite, this implies that $[\mathcal{O},r_B(b)] x = 0$ for each $x \in X$ which can only be true if $[\mathcal{O},r_B(b)] = 0$ as a map.

If we can construct a homomorphism $\ell_A: A \rightarrow B_B(X)$, then we say that $X$ is a $B$-rigged $A$-module. A $B$-rigged $A$-module can be used to induce a functor mapping between the categories of Hilbert space representations for the algebras $A$ and $B$. In general we denote the category of Hilbert space representations of a $C^*$ algebra $A$ by $\text{Mod}(A)$. At the object level, this functor takes as input a Hilbert space representation of $B$ and outputs a Hilbert space representation of $A$. Explicitly, this functor is constructed as follows. 

Let $\pi_B: B \rightarrow B(V)$ be a Hilbert space representation of $B$. Then, we can define the relative tensor product space
\beq
	X \otimes_{B} V \equiv X \otimes V/ \sim
\eeq
by quotienting the Banach space tensor product $X \otimes V$ by the equivalence relation
\beq \label{rel ten equivalence}
	x \otimes \pi_B(b) v \sim r_B(b) x \otimes v, \; \forall x \in X, b \in B, v \in V.
\eeq
This equivalence relation continues the theme that one should regard the elements of $B$ as scalars which, in \eqref{rel ten equivalence}, are allowed to move through the tensor product. To promote $X \otimes_B V$ to a Hilbert space we need to close it with respect to a pre-inner product. This is where the rigging $G_B$ comes into play. Let $g: V \times V \rightarrow \mathbb{C}$ be the inner product on $V$, then
\beq \label{preinner on rel ten}
	g_{G_B}(x_1 \otimes v_1, x_2 \otimes v_2) \equiv g(v_1, \ell_A \circ G_B(x_1,x_2) v_2)
\eeq 
is a pre-inner product on $X \otimes_B V$. Closing $X \otimes_{B} V$ with respect to the bilinear \eqref{preinner on rel ten} we obtain the Hilbert space $X \otimes_{B,G_B} V$. The representation $\ell_A$ of $X$ induces a representation $\pi_A: A \rightarrow B(X \otimes_{B,G_B} V)$ which acts as
\beq
	\pi_A(a)\bigg(x \otimes v\bigg) \equiv \ell_A(a)x \otimes v. 
\eeq
We define the Rieffel induction functor generated by the $B$-rigged $A$-module $X$ by
\beq
	F_{X}: \text{Mod}(A) \rightarrow \text{Mod}(B),
\eeq
with $F_X(V) \equiv X \otimes_{B,G_B} V$. 

A standard example of Rieffel induction arises in the special case where $B \subset A$ is a $C^*$ subalgebra. In this case, any operator valued weight $T: A \rightarrow B$ automatically renders $A$ a $B$-rigged $A$-module. Firstly, the operator valued weight can be interpreted as a $B$-valued bi-linear on $A$:
\beq
	G_T: A \times A \rightarrow B, \; G_T(a_1,a_2) = T(a_1^* a_2).
\eeq
The algebra $b$ realizes a right representation on $A$ via composition on the right, and the bi-module property of the operator valued weight ensures that
\beq
	G_T(a_1, a_2 b) = T(a_1^* a_2 b) = T(a_1^* a_2) b = G_T(a_1, a_2) b,
\eeq
which implies that $G_T$ is a $B$-valued inner product. Similarly, $A$ admits a representation acting on itself by left composition. Thus $A$ is a $B$-rigged $A$-module. Given a Hilbert space representation $V$ for the subalgebra $B$, the Rieffel induction functor $F_{T}(V) \equiv A \otimes_{B,G_T} V$ can be interpreted as a generalization of the GNS construction to operator valued weights. In particular, if $A$ is a unital $C^*$ algebra and $B = \mathbb{C} \mathbb{1}$ an operator valued weight reduces to an ordinary weight on $A$, $\varphi$. Taking $V = \mathbb{C}$ as the representation of $B$, the relative tensor product $A \otimes_{\mathbb{C}} \mathbb{C} \simeq A$ and the closure of this space is taken with respect to the pre-inner product
\beq
	g_{\varphi}(a_1, a_2) = \varphi(a_1^* a_2).
\eeq
This is precisely the pre-inner product of the GNS Hilbert space of $A$ with respect to the weight $\varphi$. 

\subsection{Imprimitivity}

Given a pair of $C^*$ algebras $A$ and $B$ an \emph{$A-B$ imprimitivity bi-module} is a $5$-tuple $\mathcal{X} \equiv (X,\ell_A, r_B, G_A, G_B)$ with $\ell_A: A \rightarrow B(X)$ a left representation, $r_B: B \rightarrow B(X)$ a right representation, $G_A: X \times X \rightarrow A$ an $A$-valued inner product and $G_B: X \times X \rightarrow B$ a $B$-valued inner product. The representations and the inner products are compatible in the sense that 
\beq
    G_A(x_1, \ell_A(a)x_2) = a G_A(x_1,x_2), \qquad G_B(x_1,r_B(b)x_2) = G_B(x_1,x_2)b, 
\eeq
and
\beq
    \ell_A \circ G_A(x_1,x_2) x_3 = r_B \circ G_B(x_2,x_3) x_1.
\eeq
Notice that, as part of the definition of $\mathcal{X}$ it indicates that $X$ is a $B$-rigged $A$-module. Thus, it may be used to induce representations of $B$ to representations of $A$. We denote the Rieffel induction functor in this case by $F_{\mathcal{X}}$. 

As we shall now demonstrate, an $A-B$ imprimitivity bi-module gives rise to a pair of adjointly related Rieffel induction functors. First, for a representation $\pi: A \rightarrow B(X)$ we define the adjoint representation $\overline{\pi}: A \rightarrow B(X)$ such that $\overline{\pi}(a) x \equiv \pi(a^*) x$. Similarly, given a bilinear $G_A: X \times X \rightarrow A$ we define the adjoint $\overline{G}_A: X \times X \rightarrow A$ by $\overline{G}_A(x_1,x_2) \equiv G_A(x,y)^* = G_A(y,x)$. It is not hard to show that, if $\ell_A$ is a left representation and $G_A$ is an $A$-valued inner product compatible with $\ell_A$, the $\overline{\ell}_A$ is a right representation compatible with the inner product $\overline{G}_A$. Thus, the $A-B$ imprimitivity bi-module $\mathcal{X}$ has a natural adjoint $\overline{\mathcal{X}} \equiv (X,\overline{r}_B, \overline{\ell}_A, \overline{G}_B, \overline{G}_A)$ which is a $B-A$ bi-module. The adjoint imprimitivity bi-module $overline{\mathcal{X}}$ realizes the space $X$ as an $A$-rigged $B$-module. Thus, it may be used to induce representations of $A$ to representations of $B$. We denote the Rieffel induction functor in this case by $F_{\overline{\mathcal{X}}}$. 

Rieffel's inversion theorem \cite{rieffel1974induced} tells us that the functors $F_{\mathcal{X}}$ and $F_{\overline{\mathcal{X}}}$ are inverses of each other. That is, for any Hilbert space representation $V$ of $B$, $F_{\overline{\mathcal{X}}} \circ F_{\mathcal{X}}(V) \simeq V$. In this sense, the existence of an $A-B$ imprimitivity bi-module implies an equivalence of the categories $\text{Mod}(A)$ and $\text{Mod}(B)$. This is called \emph{strong Morita equivalence}.\footnote{We should note, the existence of an $A-B$ imprimitivity bi-module is a sufficient but not necessary condition for strong Morita equivalence between $A$ and $B$.} The concept of strong Morita equivalence is crucial as it allows us to formulate an imprimitivity theorem identifying which representations of a given $C^*$ algebra $A$ can be regarded as having been induced from representations of the $C^*$ algebra $B$. 

Let $B$ be a $C^*$ algebra, and $(X,r_B,G_B)$ a $B$-rigged space. The key ingredient in the imprimitivity theory is the \emph{imprimitivity algebra of a $B$-rigged space}. As we shall see, the imprimitivity algebra indexes all of the possible representations which can be induced by a given $B$-rigged space. To each pair $x_1,x_2 \in X$ we assign a map $T_{x_1,x_2}: X \rightarrow X$ given by
\beq \label{T action}
    T_{x_1,x_2}(x_3) \equiv r_B(x_2,x_3) x_1. 
\eeq
The operator $T_{x_1,x_2}$ is the analog, in a $B$-rigged space, of the outer product of two vectors in a Hilbert space. The imprimitivity algebra of $X$ is defined
\beq
    E_X \equiv \{T_{x_1,x_2} \; | \; x_1,x_2 \in X\}.
\eeq
It is natural to regard the map
\beq
    G_{E_X}: X \times X \rightarrow E(X), \; (x_1,x_2) \mapsto T_{x_1,x_2}
\eeq
as an $E_X$ valued inner product on $X$. Letting $\ell_{E_X}: E_X \rightarrow B(X)$ denote the representation of $E_X$ on $X$, it is straightforward to see that 
\beq
    G_{E_X}(\ell_{E_X}(e) x_1, x_2) = e G_{E_X}(x_1,x_2),
\eeq
and by \eqref{T action}
\beq
    \ell_{E_X} \circ G_{E_X}(x_1,x_2) x_3 = r_{B}(x_2,x_3) x_1. 
\eeq
Thus, the collection $(X,\ell_{E_X},r_{B},G_{E_X},G_B)$ is a $E_X-B$ imprimitivity bi-module. By strong Morita equivalence, one may regard $E_X$ as encoding all of the possible representations which can be induced from $B$ via $X$. 

This leads us to the imprimitivity theorem: Let $A$ and $B$ be $C^*$ algebras, and suppose that $X$ is a $B$-rigged $A$-module with representations $r_B$ and $\ell_A$, respectively. Let us denote by $E_X$ the imprimitivity algebra of $X$, and by $\ell_{E_X}$ the representation of $E_X$ on $X$. Finally, let $\pi_A: A \rightarrow B(W)$ be a representation of $A$ on a Hilbert space $W$. There exists a Hilbert space representations $V$ of $B$ such that $F_X(V) \simeq W$ if and only if $W$ can be made into a Hilbert space representation of $E_X$ such that
\beq \label{Imprimitivity}
    \ell_A(a)\bigg(\ell_{E_X}(e) x\bigg) = \ell_{E_X}(ae) x, \; \forall a \in A, e \in E_X, x \in X.
\eeq
Here, $ae$ is the product of $a$ and $e$ mutually regarded as elements in $B_B(X)$.\footnote{It can be shown that $E_X$ is a two-sided ideal in $B_B(X)$, hence why the product $ae \in E_X$.} 

\subsection{Rieffel Induction as a Quotienting}

It has been argued that Rieffel induction should be interpreted as an algebraic analog of symplectic reduction \cite{landsman1995rieffel}. Let $H$ be a locally compact group and $\mathcal{L}(H)$ its associated group von Neumann algebra. In particular, the space of bounded operators on a $\mathcal{L}(H)$-rigged space $X$ with a definite $\mathcal{L}(H)$ valued inner product is the analog of the classical constraint surface. Recall that bounded operators $B_{\mathcal{L}(H)}(X)$ are adjointable, bounded linear maps $\mathcal{O}: X \rightarrow X$ and automatically commute with the right representation of $H$ on $X$. Specifying any Hilbert space representation of $\mathcal{L}(H)$, or equivalently any unitary representation of $H$, we can induce representations $\ell_A: A \rightarrow B_{\mathcal{L}(H)}(X)$ to Hilbert space representations commuting with constraints encoded by the representation $r_{\mathcal{L}(H)}: \mathcal{L}(H) \rightarrow B(X)$. In other words, $B_{\mathcal{L}(H)}(X)$ may be interpreted as playing the role of a $H$-framed algebra with its various subalgebras coinciding with crossed product charts. 

It is straightforward to construct $\mathcal{L}(H)$-rigged spaces. Suppose that $X$ is a Hilbert space with inner product $g_X$, $H$ is a locally compact group, and $X$ admits a unitary representation $U: H \rightarrow U(X)$. Then, we have a natural rigging:
\beq
	G^{\mathcal{L}(H)}: X \times X \rightarrow \mathcal{L}(H), \; G^{\mathcal{L}(H)}_{x_1,x_2}(h) \equiv \delta(h)^{1/2} g_K(x_1, U(h) x_2). 
\eeq
A generic element in $\mathcal{L}(G)$ can be regarded as a compactly supported map $\psi: G \rightarrow \mathbb{C}$ and acts on $X$ via the (right) representation:
\beq
	r_{\mathcal{L}(H)}(\psi) = \int_{H} \mu(h) \; \psi(h) U(h^{-1}).
\eeq
It isn't hard to show that
\beq	
	G^{\mathcal{L}(H)}_{x_1, r_{\mathcal{L}(H)}(\psi) x_2}(h) = \bigg(G^{\mathcal{L}(H)}_{x_1,x_2} \star \psi\bigg)(h),
\eeq
where $\star$ is the convolutional product in $\mathcal{L}(H)$. 

It is tempting to interpret the algebra $B_{\mathcal{L}(H)}(X)$ associated with a $\mathcal{L}(H)$-rigged space $X$ as encoding a QRF, viewed here as a quotient algebra relative to a specified action of $H$. The set of algebras $A$ with representations that may be induced from $X$, as identified from the imprimitivity theorem via $E_X$, could then coincide with local charts refined by (i.e. contained within) $B_{\mathcal{L}(H)}(X)$. In this way, it seems reasonable to expect that a $G$-framed algebra may be constructed as the union of algebras $B_{\mathcal{L}(H)}(X)$ for various choices of $X$.

\bibliographystyle{uiuchept}
\bibliography{qrf}
\end{document}

%% file: packages.tex
\usepackage{amsmath,amssymb,amsfonts,bbm,bm}
\usepackage{slashed}

\usepackage{tocloft} 

\usepackage[nosort]{cite} 
\usepackage{graphicx,color}
\usepackage[colorlinks=true, linkcolor=blue, citecolor=blue, linktoc=all]{hyperref}

\usepackage[usenames,dvipsnames]{xcolor}

\usepackage{tikz-cd}
\usepackage{pict2e}
\usepackage{physics}
\usepackage{comment} 
\usepackage[makeroom]{cancel}
\usepackage{musicography}

%% file: standard.tex
\newcommand{\beq}{\begin{eqnarray}}
\newcommand{\eeq}{\end{eqnarray}}
\newcommand{\beqn}{\begin{eqnarray}}
\newcommand{\eeqn}{\end{eqnarray}}
\newcommand{\bea}{\begin{eqnarray}}
\newcommand{\eea}{\end{eqnarray}}
\newcommand{\be}{\begin{equation}}
\newcommand{\ee}{\end{equation}}

\newcommand{\ack}[1]{[{\bf Pfft!: {#1}}]}

\newcommand{\un}[1]{\underline{#1}}

\usepackage[normalem]{ulem}

\newcommand{\uiuc}[1]{
	\centerline{
		\begin{minipage}[c]{0.7\textwidth}
			\begin{center}
			${}^{#1}$ Illinois Center for Advanced Studies of the Universe \& Department of Physics,\\ 
			University of Illinois, 1110 West Green St., Urbana IL 61801, U.S.A.
			\end{center}
		\end{minipage}
		}
	}
 \newcommand{\nyu}[1]{
	\centerline{
		\begin{minipage}[c]{0.7\textwidth}
			\begin{center}
			${}^{#1}$ Center for Cosmology and Particle Physics, New York University, New York, NY 10003, USA
			\end{center}
		\end{minipage}
		}
	}
  \newcommand{\upenn}[1]{
	\centerline{
		\begin{minipage}[c]{0.7\textwidth}
			\begin{center}
			${}^{#1}$David Rittenhouse Laboratory, University of Pennsylvania, Philadelphia, PA 19104, USA
			\end{center}
		\end{minipage}
		}
	}

%% file: Alg_Macro.tex
\usepackage{mathrsfs,mathtools}
\renewcommand\mathbb[1]{\mathbbm{#1}}

\newcommand\saa[1]{{\color{teal}[\textbf{Shadi}: #1]}}
\newcommand\MK[1]{{\textcolor{red}{MK: #1}}}

\makeatletter
\DeclareRobustCommand{\loplus}{\mathbin{\mathpalette\dog@lsemi{+}}}
\DeclareRobustCommand{\lotimes}{\mathbin{\mathpalette\dog@lsemi{\times}}}
\DeclareRobustCommand{\roplus}{\mathbin{\mathpalette\dog@rsemi{+}}}
\DeclareRobustCommand{\rotimes}{\mathbin{\mathpalette\dog@rsemi{\times}}}

\newcommand{\dog@rsemi}[2]{\dog@semi{#1}{#2}{-90,90}}
\newcommand{\dog@lsemi}[2]{\dog@semi{#1}{#2}{270,90}}
\newcommand{\dog@semi}[3]{%
  \begingroup
  \sbox\z@{$\m@th#1#2$}%
  \setlength{\unitlength}{\dimexpr\ht\z@+\dp\z@\relax}%
  \makebox[\wd\z@]{\raisebox{-\dp\z@}{%
    \begin{picture}(1,1)
    \linethickness{\variable@rule{#1}}
    \roundcap
    \put(0.5,0.5){\makebox(0,0){\raisebox{\dp\z@}{$\m@th#1#2$}}}
    \put(0.5,0.5){\arc[#3]{0.5}}
    \end{picture}%
  }}%
  \endgroup
}
\newcommand{\variable@rule}[1]{%
  \fontdimen8  
  \ifx#1\displaystyle\textfont3\else
    \ifx#1\textstyle\textfont3\else
      \ifx#1\scriptstyle\scriptfont3\else
        \scriptscriptfont3\relax
  \fi\fi\fi
}
\DeclareRobustCommand{\loplus}{\mathbin{\mathpalette\dog@lsemi{+}}}

%% file: qrf.bbl
\providecommand{\noopsort}[1]{}\providecommand{\singleletter}[1]{#1}%
\providecommand{\href}[2]{#2}\begingroup\raggedright\begin{thebibliography}{10}

\bibitem{dirac1964lectures}
P.~Dirac, {\em Lectures on Quantum Mechanics}.
\newblock Belfer Graduate School of Science. Monographs series. Yeshiva
  University, 1964.

\bibitem{dirac_lectures_2013}
P.~A.~M. Dirac, {\em Lectures on quantum mechanics}.
\newblock Snowball, 2013.

\bibitem{Ashtekar:1995zh}
A.~Ashtekar, J.~Lewandowski, D.~Marolf, J.~Mourao, and T.~Thiemann,
  ``{Quantization of diffeomorphism invariant theories of connections with
  local degrees of freedom},'' \href{http://dx.doi.org/10.1063/1.531252}{{\em
  J. Math. Phys.} {\bf 36} (1995)  6456--6493},
  \href{http://arxiv.org/abs/gr-qc/9504018}{{\tt arXiv:gr-qc/9504018}}.

\bibitem{Marolf:1996gb}
D.~Marolf, ``{Path integrals and instantons in quantum gravity: Minisuperspace
  models},'' \href{http://dx.doi.org/10.1103/PhysRevD.53.6979}{{\em Phys. Rev.
  D} {\bf 53} (1996)  6979--6990},
  \href{http://arxiv.org/abs/gr-qc/9602019}{{\tt arXiv:gr-qc/9602019}}.

\bibitem{Giulini:1998kf}
D.~Giulini and D.~Marolf, ``{A Uniqueness theorem for constraint
  quantization},'' \href{http://dx.doi.org/10.1088/0264-9381/16/7/322}{{\em
  Class. Quant. Grav.} {\bf 16} (1999)  2489--2505},
  \href{http://arxiv.org/abs/gr-qc/9902045}{{\tt arXiv:gr-qc/9902045}}.

\bibitem{Giulini:1998rk}
D.~Giulini and D.~Marolf, ``{On the generality of refined algebraic
  quantization},'' \href{http://dx.doi.org/10.1088/0264-9381/16/7/321}{{\em
  Class. Quant. Grav.} {\bf 16} (1999)  2479--2488},
  \href{http://arxiv.org/abs/gr-qc/9812024}{{\tt arXiv:gr-qc/9812024}}.

\bibitem{Marolf:2000iq}
D.~Marolf, ``{Group averaging and refined algebraic quantization: Where are we
  now?},'' in {\em {9th Marcel Grossmann Meeting on Recent Developments in
  Theoretical and Experimental General Relativity, Gravitation and Relativistic
  Field Theories (MG 9)}}.
\newblock 7, 2000.
\newblock \href{http://arxiv.org/abs/gr-qc/0011112}{{\tt arXiv:gr-qc/0011112}}.

\bibitem{Ashtekar1991-ASHCPO}
A.~Ashtekar and J.~Stachel, eds., {\em Conceptual Problems of Quantum Gravity}.
\newblock Birkhauser, 1991.

\bibitem{rovelli_2004}
C.~Rovelli, \href{http://dx.doi.org/10.1017/CBO9780511755804}{{\em Quantum
  Gravity}}.
\newblock Cambridge Monographs on Mathematical Physics. Cambridge University
  Press, 2004.

\bibitem{thiemann_2007}
T.~Thiemann, \href{http://dx.doi.org/10.1017/CBO9780511755682}{{\em Modern
  Canonical Quantum General Relativity}}.
\newblock Cambridge Monographs on Mathematical Physics. Cambridge University
  Press, 2007.

\bibitem{Witten:2023xze}
E.~Witten, ``{A background-independent algebra in quantum gravity},''
  \href{http://dx.doi.org/10.1007/JHEP03(2024)077}{{\em JHEP} {\bf 03} (2024)
  077}, \href{http://arxiv.org/abs/2308.03663}{{\tt arXiv:2308.03663
  [hep-th]}}.

\bibitem{aharonov_quantum_1984}
Y.~Aharonov and T.~Kaufherr, ``{{Quantum Frames of Reference}},''
  \href{http://dx.doi.org/10.1103/PhysRevD.30.368}{{\em Phys. Rev. D} {\bf 30}
  (1984) no.~2, 368}.

\bibitem{rovelli_quantum_1991}
C.~Rovelli, ``{Quantum reference systems},''
  \href{http://dx.doi.org/10.1088/0264-9381/8/2/012}{{\em Class. Qu. Grav.}
  {\bf 8} (1991) no.~2, 317}.

\bibitem{kitaev_superselection_2004}
A.~Kitaev, D.~Mayers, and J.~Preskill, ``Superselection rules and quantum
  protocols,'' \href{http://dx.doi.org/10.1103/PhysRevA.69.052326}{{\em Phys.
  Rev. A} {\bf 69} (2004) no.~5, 052326}.

\bibitem{bartlett_reference_2007}
S.~D. Bartlett, T.~Rudolph, and R.~W. Spekkens, ``Reference frames,
  superselection rules, and quantum information,''
  \href{http://dx.doi.org/10.1103/RevModPhys.79.555}{{\em Rev. Mod. Phys.} {\bf
  79} (2007) no.~2, 555--609}.

\bibitem{gour_resource_2008}
G.~Gour and R.~W. Spekkens, ``The resource theory of quantum reference frames:
  manipulations and monotones,''
  \href{http://dx.doi.org/10.1088/1367-2630/10/3/033023}{{\em New Journal of
  Physics} {\bf 10} (2008) no.~3, 033023}.

\bibitem{girelli_quantum_2008}
F.~Girelli and D.~Poulin, ``Quantum reference frames and deformed symmetries,''
  \href{http://dx.doi.org/10.1103/PhysRevD.77.104012}{{\em Phys. Rev. D} {\bf
  77} (2008) no.~10, 104012}. Publisher: American Physical Society.

\bibitem{bartlett_quantum_2009}
S.~D. Bartlett, T.~Rudolph, R.~W. Spekkens, and P.~S. Turner, ``Quantum
  communication using a bounded-size quantum reference frame,''
  \href{http://dx.doi.org/10.1088/1367-2630/11/6/063013}{{\em New Journal of
  Physics} {\bf 11} (2009) no.~6, 063013}.

\bibitem{angelo_physics_2011}
R.~M. Angelo, N.~Brunner, S.~Popescu, A.~J. Short, and P.~Skrzypczyk, ``Physics
  within a quantum reference frame,''
  \href{http://dx.doi.org/10.1088/1751-8113/44/14/145304}{{\em J. Phys. A} {\bf
  44} (2011) no.~14, 145304}.

\bibitem{angelo_kinematics_2012}
R.~M. Angelo and A.~D. Ribeiro, ``{Kinematics and dynamics in noninertial
  quantum frames of reference},''
  \href{http://dx.doi.org/10.1088/1751-8113/45/46/465306}{{\em J. Phys. A} {\bf
  45} (2012) no.~46, 465306}.

\bibitem{palmer_changing_2014}
M.~C. Palmer, F.~Girelli, and S.~D. Bartlett, ``Changing quantum reference
  frames,'' \href{http://dx.doi.org/10.1103/PhysRevA.89.052121}{{\em Physical
  Review A} {\bf 89} (2014) no.~5, 052121}.
  \url{https://link.aps.org/doi/10.1103/PhysRevA.89.052121}. Publisher:
  American Physical Society.

\bibitem{pienaar_relational_2016}
J.~Pienaar, ``{A relational approach to quantum reference frames for spins},''
  \href{http://arxiv.org/abs/1601.07320}{{\tt arXiv:1601.07320 [quant-ph]}}.

\bibitem{smith_quantum_2016}
A.~R.~H. Smith, M.~Piani, and R.~B. Mann, ``{Quantum reference frames
  associated with noncompact groups: The case of translations and boosts and
  the role of mass},'' \href{http://dx.doi.org/10.1103/physreva.94.012333}{{\em
  Phys. Rev. A} {\bf 94} (2016) no.~1, }.

\bibitem{miyadera_approximating_2016}
P.~Busch, L.~Loveridge, and T.~Miyadera, ``{Approximating relational
  observables by absolute quantities: a quantum accuracy-size trade-off},''
  \href{http://dx.doi.org/10.1088/1751-8113/49/18/185301}{{\em J. Phys. A} {\bf
  49} (2016) no.~18, 185301}.

\bibitem{loveridge_relativity_2017}
L.~Loveridge, P.~Busch, and T.~Miyadera, ``{Relativity of quantum states and
  observables},'' \href{http://dx.doi.org/10.1209/0295-5075/117/40004}{{\em
  EPL} {\bf 117} (2017) no.~4, 40004},
  \href{http://arxiv.org/abs/1604.02836}{{\tt arXiv:1604.02836 [quant-ph]}}.

\bibitem{belenchia_quantum_2018}
A.~Belenchia, R.~M. Wald, F.~Giacomini, E.~Castro-Ruiz, v.~Brukner, and
  M.~Aspelmeyer, ``{Quantum Superposition of Massive Objects and the
  Quantization of Gravity},''
  \href{http://dx.doi.org/10.1103/PhysRevD.98.126009}{{\em Phys. Rev. D} {\bf
  98} (2018) no.~12, 126009}, \href{http://arxiv.org/abs/1807.07015}{{\tt
  arXiv:1807.07015 [quant-ph]}}.

\bibitem{giacomini_quantum_2019}
F.~Giacomini, E.~Castro-Ruiz, and v.~Brukner, ``{Quantum mechanics and the
  covariance of physical laws in quantum reference frames},''
  \href{http://dx.doi.org/10.1038/s41467-018-08155-0}{{\em Nature Commun.} {\bf
  10} (2019) no.~1, 494}, \href{http://arxiv.org/abs/1712.07207}{{\tt
  arXiv:1712.07207 [quant-ph]}}.

\bibitem{hoehn_equivalence_2020}
P.~A. Hoehn, A.~R.~H. Smith, and M.~P.~E. Lock, ``{Equivalence of Approaches to
  Relational Quantum Dynamics in Relativistic Settings},''
  \href{http://dx.doi.org/10.3389/fphy.2021.587083}{{\em Front. in Phys.} {\bf
  9} (2021)  181}, \href{http://arxiv.org/abs/2007.00580}{{\tt arXiv:2007.00580
  [gr-qc]}}.

\bibitem{castro-ruiz_quantum_2020}
E.~Castro-Ruiz, F.~Giacomini, A.~Belenchia, and v.~Brukner, ``{Quantum clocks
  and the temporal localisability of events in the presence of gravitating
  quantum systems},'' \href{http://dx.doi.org/10.1038/s41467-020-16013-1}{{\em
  Nature Commun.} {\bf 11} (2020) no.~1, 2672},
  \href{http://arxiv.org/abs/1908.10165}{{\tt arXiv:1908.10165 [quant-ph]}}.

\bibitem{Danielson:2021egj}
D.~L. Danielson, G.~Satishchandran, and R.~M. Wald, ``{Gravitationally mediated
  entanglement: Newtonian field versus gravitons},''
  \href{http://dx.doi.org/10.1103/PhysRevD.105.086001}{{\em Phys. Rev. D} {\bf
  105} (2022) no.~8, 086001}, \href{http://arxiv.org/abs/2112.10798}{{\tt
  arXiv:2112.10798 [quant-ph]}}.

\bibitem{hoehn_trinity_2021}
P.~A. Hoehn, A.~R.~H. Smith, and M.~P.~E. Lock, ``{Trinity of relational
  quantum dynamics},''
  \href{http://dx.doi.org/10.1103/PhysRevD.104.066001}{{\em Phys. Rev. D} {\bf
  104} (2021) no.~6, 066001}, \href{http://arxiv.org/abs/1912.00033}{{\tt
  arXiv:1912.00033 [quant-ph]}}.

\bibitem{castro-ruiz_relative_2021}
E.~Castro-Ruiz and O.~Oreshkov, ``{Relative subsystems and quantum reference
  frame transformations},'' \href{http://arxiv.org/abs/2110.13199}{{\tt
  arXiv:2110.13199 [quant-ph]}}.

\bibitem{giacomini_spacetime_2021}
F.~Giacomini, ``{Spacetime Quantum Reference Frames and superpositions of
  proper times},'' \href{http://dx.doi.org/10.22331/q-2021-07-22-508}{{\em
  Quantum} {\bf 5} (2021)  508}, \href{http://arxiv.org/abs/2101.11628}{{\tt
  arXiv:2101.11628 [quant-ph]}}.

\bibitem{ali_ahmad_quantum_2022}
S.~Ali~Ahmad, T.~D. Galley, P.~A. Hohn, M.~P. Lock, and A.~R. Smith, ``{Quantum
  Relativity of Subsystems},''
  \href{http://dx.doi.org/10.1103/PhysRevLett.128.170401}{{\em Phys. Rev.
  Lett.} {\bf 128} (2022) no.~17, 170401}.

\bibitem{cepollaro_quantum_2022}
C.~Cepollaro and F.~Giacomini, ``{Quantum generalisation of Einstein's
  Equivalence Principle can be verified with entangled clocks as quantum
  reference frames},'' \href{http://arxiv.org/abs/2112.03303}{{\tt
  arXiv:2112.03303 [quant-ph]}}.

\bibitem{giacomini_second-quantized_2022}
F.~Giacomini and A.~Kempf, ``{Second-quantized Unruh-DeWitt detectors and their
  quantum reference frame transformations},''
  \href{http://dx.doi.org/10.1103/PhysRevD.105.125001}{{\em Phys. Rev. D} {\bf
  105} (2022) no.~12, 125001}, \href{http://arxiv.org/abs/2201.03120}{{\tt
  arXiv:2201.03120 [quant-ph]}}.

\bibitem{apadula_quantum_2022}
L.~Apadula, E.~Castro-Ruiz, and v.~Brukner, ``{Quantum Reference Frames for
  Lorentz Symmetry},'' \href{http://arxiv.org/abs/2212.14081}{{\tt
  arXiv:2212.14081 [quant-ph]}}.

\bibitem{witten_gravity_2022}
E.~Witten, ``{Gravity and the crossed product},''
  \href{http://dx.doi.org/10.1007/JHEP10(2022)008}{{\em JHEP} {\bf 10} (2022)
  008}, \href{http://arxiv.org/abs/2112.12828}{{\tt arXiv:2112.12828
  [hep-th]}}.

\bibitem{chandrasekaran_large_2022}
V.~Chandrasekaran, G.~Penington, and E.~Witten, ``{Large N algebras and
  generalized entropy},'' \href{http://dx.doi.org/10.1007/JHEP04(2023)009}{{\em
  JHEP} {\bf 04} (2023)  009}, \href{http://arxiv.org/abs/2209.10454}{{\tt
  arXiv:2209.10454 [hep-th]}}.

\bibitem{chandrasekaran_algebra_2022}
V.~Chandrasekaran, R.~Longo, G.~Penington, and E.~Witten, ``{An algebra of
  observables for de Sitter space},''
  \href{http://dx.doi.org/10.1007/JHEP02(2023)082}{{\em JHEP} {\bf 02} (2023)
  082}, \href{http://arxiv.org/abs/2206.10780}{{\tt arXiv:2206.10780
  [hep-th]}}.

\bibitem{AliAhmad:2023etg}
S.~Ali~Ahmad and R.~Jefferson, ``{Crossed product algebras and generalized
  entropy for subregions},''
  \href{http://dx.doi.org/10.21468/SciPostPhysCore.7.2.020}{{\em SciPost Phys.
  Core} {\bf 7} (2024)  020}, \href{http://arxiv.org/abs/2306.07323}{{\tt
  arXiv:2306.07323 [hep-th]}}.

\bibitem{Klinger:2023tgi}
M.~S. Klinger and R.~G. Leigh, ``{Crossed Products, Extended Phase Spaces and
  the Resolution of Entanglement Singularities},''
  \href{http://dx.doi.org/10.1016/j.nuclphysb.2024.116453}{{\em Nucl. Phys. B}
  {\bf 999} (2024)  116453}, \href{http://arxiv.org/abs/2306.09314}{{\tt
  arXiv:2306.09314 [hep-th]}}.

\bibitem{Klinger:2023auu}
M.~S. Klinger and R.~G. Leigh, ``{Crossed Products, Conditional Expectations
  and Constraint Quantization},'' \href{http://arxiv.org/abs/2312.16678}{{\tt
  arXiv:2312.16678 [hep-th]}}.

\bibitem{van1978continuous}
A.~Van~Daele, {\em Continuous crossed products and type III von Neumann
  algebras}, vol.~31.
\newblock Cambridge University Press, 1978.

\bibitem{page_evolution_1983}
D.~N. Page and W.~K. Wootters, ``{Evolution without evolution: {Dynamics}
  described by stationary observables},''
  \href{http://dx.doi.org/10.1103/PhysRevD.27.2885}{{\em Phys. Rev. D} {\bf 27}
  (1983)  2885}.

\bibitem{wootters_time_1984}
W.~K. Wootters, ``{\textquotedblleft{}Time\textquotedblright{} replaced by
  quantum correlations},'' \href{http://dx.doi.org/10.1007/BF02214098}{{\em
  Int. J. Theor. Phys.} {\bf 23} (1984) no.~8, 701--711}.

\bibitem{landstad1979duality}
M.~B. Landstad, ``Duality theory for covariant systems,'' {\em Trans. Am. Math.
  Soc.} {\bf 248} (1979) no.~2, 223--267.

\bibitem{AliAhmad:2021adn}
S.~Ali~Ahmad, T.~D. Galley, P.~A. Hoehn, M.~P.~E. Lock, and A.~R.~H. Smith,
  ``{Quantum Relativity of Subsystems},''
  \href{http://dx.doi.org/10.1103/PhysRevLett.128.170401}{{\em Phys. Rev.
  Lett.} {\bf 128} (2022) no.~17, 170401},
  \href{http://arxiv.org/abs/2103.01232}{{\tt arXiv:2103.01232 [quant-ph]}}.

\bibitem{delaHamette:2021oex}
A.-C. de~la Hamette, T.~D. Galley, P.~A. Hoehn, L.~Loveridge, and M.~P.
  Mueller, ``{Perspective-neutral approach to quantum frame covariance for
  general symmetry groups},'' \href{http://arxiv.org/abs/2110.13824}{{\tt
  arXiv:2110.13824 [quant-ph]}}.

\bibitem{Hoehn:2023sub}
P.~A. Hoehn, I.~Kotecha, and F.~M. Mele, ``{Quantum Frame Relativity of
  Subsystems, Correlations and Thermodynamics},''
  \href{http://arxiv.org/abs/2308.09131}{{\tt arXiv:2308.09131 [quant-ph]}}.

\bibitem{caramello2022introduction}
F.~C. Caramello, ``Introduction to orbifolds,''
  \href{http://arxiv.org/abs/math.DG/1909.08699}{{\tt
  arXiv:math.DG/1909.08699}}.

\bibitem{Langmann:1993pb}
E.~Langmann and G.~W. Semenoff, ``{Gribov ambiguity and nontrivial vacuum
  structure of gauge theories on a cylinder},''
  \href{http://dx.doi.org/10.1016/0370-2693(93)91436-Q}{{\em Phys. Lett. B}
  {\bf 303} (1993)  303--307}, \href{http://arxiv.org/abs/hep-th/9212038}{{\tt
  arXiv:hep-th/9212038}}.

\bibitem{1994CMaPh.160..431M}
D.~{McMullan}, ``{Constrained quantisation, gauge fixing and the Gribov
  ambiguity},'' \href{http://dx.doi.org/10.1007/BF02173423}{{\em Comm. Math.
  Phys.} {\bf 160} (1994) no.~3, 431}.

\bibitem{Fewster:2024pur}
C.~J. Fewster, D.~W. Janssen, L.~D. Loveridge, K.~Rejzner, and J.~Waldron,
  ``{Quantum reference frames, measurement schemes and the type of local
  algebras in quantum field theory},''
  \href{http://arxiv.org/abs/2403.11973}{{\tt arXiv:2403.11973 [math-ph]}}.

\bibitem{DeVuyst:2024pop}
J.~De~Vuyst, S.~Eccles, P.~A. Hoehn, and J.~Kirklin, ``{Gravitational entropy
  is observer-dependent},'' \href{http://arxiv.org/abs/2405.00114}{{\tt
  arXiv:2405.00114 [hep-th]}}.

\bibitem{Gomez:2022eui}
C.~Gomez, ``{Cosmology as a Crossed Product},''
  \href{http://arxiv.org/abs/2207.06704}{{\tt arXiv:2207.06704 [hep-th]}}.

\bibitem{Gomez:2023wrq}
C.~Gomez, ``{Entanglement, Observers and Cosmology: a view from von Neumann
  Algebras},'' \href{http://arxiv.org/abs/2302.14747}{{\tt arXiv:2302.14747
  [hep-th]}}.

\bibitem{Gomez:2023upk}
C.~Gomez, ``{Clocks, Algebras and Cosmology},''
  \href{http://arxiv.org/abs/2304.11845}{{\tt arXiv:2304.11845 [hep-th]}}.

\bibitem{haagerup1978dualI}
U.~Haagerup, ``On the dual weights for crossed products of von neumann algebras
  i: removing separability conditions,'' {\em Math. Scand.} {\bf 43} (1978)
  no.~1, 99--118.

\bibitem{haagerup1978dualII}
U.~Haagerup, ``On the dual weights for crossed products of von neumann algebras
  ii: application of operator valued weights,'' {\em Math. Scand.} {\bf 43}
  (1978) no.~1, 119--140.

\bibitem{Takesaki2}
M.~Takesaki, {\em {Theory of Operator Algebras II}}.
\newblock Springer, 2003.

\bibitem{takesaki1973duality}
M.~Takesaki, ``{Duality for crossed products and the structure of von Neumann
  algebras of type III},'' \href{http://dx.doi.org/10.1007/BF02392041}{{\em
  Acta Math.} {\bf 131} (1973)  249}.

\bibitem{doplicher1966covariance}
S.~Doplicher, D.~Kastler, and D.~W. Robinson, ``Covariance algebras in field
  theory and statistical mechanics,'' {\em Comm. Math. Phys.} {\bf 3} (1966)
  no.~1, 1--28.

\bibitem{Page:1989br}
D.~N. Page, ``{Time as an Inaccessible Observable},''.

\bibitem{vanrietvelde_switching_2021}
A.~Vanrietvelde, P.~A. Hoehn, and F.~Giacomini, ``{Switching quantum reference
  frames in the N-body problem and the absence of global relational
  perspectives},'' \href{http://dx.doi.org/10.22331/q-2023-08-22-1088}{{\em
  Quantum} {\bf 7} (2023)  1088}, \href{http://arxiv.org/abs/1809.05093}{{\tt
  arXiv:1809.05093 [quant-ph]}}.

\bibitem{Sorce:2023fdx}
J.~Sorce, ``{Notes on the Type Classification of von Neumann Algebras},''
  \href{http://dx.doi.org/10.1142/S0129055X24300024}{{\em Rev. Math. Phys.}
  {\bf 36} (2024) no.~02, 2430002}, \href{http://arxiv.org/abs/2302.01958}{{\tt
  arXiv:2302.01958 [hep-th]}}.

\bibitem{Klinger:2023qna}
M.~S. Klinger, R.~G. Leigh, and P.-C. Pai, ``{Extended phase space in general
  gauge theories},''
  \href{http://dx.doi.org/10.1016/j.nuclphysb.2023.116404}{{\em Nucl. Phys. B}
  {\bf 998} (2024)  116404}, \href{http://arxiv.org/abs/2303.06786}{{\tt
  arXiv:2303.06786 [hep-th]}}.

\bibitem{berkson_representations_1969}
E.~Berkson and H.~Porta, ``{Representations of B(X)},''
  \href{http://dx.doi.org/10.1016/0022-1236(69)90048-2}{{\em J. Func. Anal.}
  {\bf 3} (1969) no.~1, 1--34}.

\bibitem{khodsiani2015type}
B.~Khodsiani and A.~Rejali, ``A type of gns-construction for banach algebras,''
  \href{http://arxiv.org/abs/1504.07069}{{\tt arXiv:1504.07069 [math.RT]}}.

\bibitem{hochschild1947cohomology}
G.~Hochschild, ``Cohomology and representations of associative algebras,''.

\bibitem{kadison1971cohomology}
R.~V. Kadison and J.~R. Ringrose, ``Cohomology of operator algebras: I. type i
  von neumann algebras,''.

\bibitem{raeburn2000twisted}
I.~Raeburn, A.~Sims, and D.~P. Williams, ``Twisted actions and obstructions in
  group cohomology,'' in {\em C*-Algebras: Proceedings of the SFB-Workshop on
  C*-Algebras, M{\"u}nster, Germany, March 8--12, 1999}, pp.~161--181,
  Springer.
\newblock 2000.

\bibitem{sutherland1980cohomology}
C.~E. Sutherland, ``Cohomology and extensions of von neumann algebras. i,''
  {\em Publications of the Research Institute for Mathematical Sciences} {\bf
  16} (1980) no.~1, 105--133.

\bibitem{sutherland1980cohomologyII}
C.~E. Sutherland, ``Cohomology and extensions of von neumann algebras. ii,''
  {\em Publications of the Research Institute for Mathematical Sciences} {\bf
  16} (1980) no.~1, 135--174.

\bibitem{landstad1987representations}
M.~B. Landstad, J.~Phillips, I.~Raeburn, and C.~E. Sutherland,
  ``Representations of crossed products by coactions and principal bundles,''
  {\em Trans. Am. Math. Soc.} {\bf 299} (1987) no.~2, 747--784.

\bibitem{Carrozza_2022}
S.~Carrozza and P.~A. Hoehn, ``{Edge modes as reference frames and boundary
  actions from post-selection},''
  \href{http://dx.doi.org/10.1007/JHEP02(2022)172}{{\em JHEP} {\bf 02} (2022)
  172}, \href{http://arxiv.org/abs/2109.06184}{{\tt arXiv:2109.06184
  [hep-th]}}.

\bibitem{carrozza2023edgemodesdynamicalframes}
S.~Carrozza, S.~Eccles, and P.~A. Hoehn, ``{Edge modes as dynamical frames:
  charges from post-selection in generally covariant theories},''
  \href{http://arxiv.org/abs/2205.00913}{{\tt arXiv:2205.00913 [hep-th]}}.

\bibitem{goeller2022diffeomorphisminvariantobservablesdynamicalframes}
C.~Goeller, P.~A. Hoehn, and J.~Kirklin, ``{Diffeomorphism-invariant
  observables and dynamical frames in gravity: reconciling bulk locality with
  general covariance},'' \href{http://arxiv.org/abs/2206.01193}{{\tt
  arXiv:2206.01193 [hep-th]}}.

\bibitem{Kabel_2023}
V.~Kabel, v.~Brukner, and W.~Wieland, ``{Quantum reference frames at the
  boundary of spacetime},''
  \href{http://dx.doi.org/10.1103/PhysRevD.108.106022}{{\em Phys. Rev. D} {\bf
  108} (2023) no.~10, 106022}, \href{http://arxiv.org/abs/2302.11629}{{\tt
  arXiv:2302.11629 [gr-qc]}}.

\bibitem{Ciambelli:2021ujl}
L.~Ciambelli and R.~G. Leigh, ``{Lie algebroids and the geometry of off-shell
  BRST},'' \href{http://dx.doi.org/10.1016/j.nuclphysb.2021.115553}{{\em Nucl.
  Phys. B} {\bf 972} (2021)  115553},
  \href{http://arxiv.org/abs/2101.03974}{{\tt arXiv:2101.03974 [hep-th]}}.

\bibitem{Jia:2023tki}
W.~Jia, M.~S. Klinger, and R.~G. Leigh, ``{BRST cohomology is Lie algebroid
  cohomology},'' \href{http://dx.doi.org/10.1016/j.nuclphysb.2023.116317}{{\em
  Nucl. Phys. B} {\bf 994} (2023)  116317},
  \href{http://arxiv.org/abs/2303.05540}{{\tt arXiv:2303.05540 [hep-th]}}.

\bibitem{rieffel1974induced}
M.~A. Rieffel, ``Induced representations of c*-algebras,'' {\em Adv. Math.}
  {\bf 13} (1974) no.~2, 176--257.

\bibitem{landsman1995rieffel}
N.~P. Landsman, ``{Rieffel Induction as generalized quantum Marsden-Weinstein
  reduction},'' {\em J. Geom. Phys.} {\bf 15} (1995) no.~4, 285.

\bibitem{marsden1974reduction}
J.~Marsden and A.~Weinstein, ``Reduction of symplectic manifolds with
  symmetry,'' {\em Rep. Math. Phys.} {\bf 5} (1974) no.~1, 121--130.

\end{thebibliography}\endgroup
